\documentclass[12pt,notoc]{JHEP}

\usepackage{amsmath,amssymb,euscript,array,cite}

\input{epsf}
\newcommand{\startappendix}{
\setcounter{section}{0}
\renewcommand{\thesection}{\Alph{section}}}
\newcommand{\Appendix}[1]{
\refstepcounter{section}
\begin{flushleft}
{\large\bf Appendix \thesection: #1}
\end{flushleft}}
\usepackage{epsfig}

\def\htau{\tau}

\def\N{{\cal N}}

\def\ttau{{\tilde\tau}}

\def\Tr{{\rm Tr}}
\def\sst{\scriptscriptstyle}

\def\C{{\bf C}}
\def\Z{{\bf Z}}
\def\X{{\bf X}}
\def\RR{{\mathfrak R}}

\def\CC{{\cal C}}
\def\Dbarslash{\,\,{\raise.15ex\hbox{/}\mkern-12mu {\bar\D}}}
\def\Dslash{\,\,{\raise.15ex\hbox{/}\mkern-12mu \D}}
\def\delslash{\,\,{\raise.15ex\hbox{/}\mkern-9mu \partial}}
\def\delbarslash{\,\,{\raise.15ex\hbox{/}\mkern-9mu {\bar\partial}}}

\def\BB{{\EuScript B}}

\def\AA{{\EuScript A}}

\newcommand{\MAT}[1]{\begin{pmatrix} #1\end{pmatrix}}
\newcommand{\EQ}[1]{\begin{equation} #1 \end{equation}}
\newcommand{\AL}[1]{\begin{subequations}\begin{align} #1
\end{align}\end{subequations}}
\newcommand{\SP}[1]{\begin{equation}\begin{split} #1
\end{split}\end{equation}}

\title{An $\N=1$ duality cascade from a deformation of $\N=4$ SUSY
Yang-Mills theory }
\author{Timothy J.~Hollowood ${}^a$ and S. Prem Kumar ${}^{a,b}$\\

${}^a$ Department of Physics,\\ University of Wales Swansea,\\
Swansea, SA2 8PP, UK.\\
\vspace{0.2 in}
${}^b$ DAMTP,\\
University of Cambridge,\\
Wilberforce Road,\\
Cambridge, CB3 0WA, UK.\\
\vspace{0.2 in}
E-mail: {\tt t.hollowood@swan.ac.uk},
{\tt s.p.kumar@damtp.cam.ac.uk}}
\preprint{DAMTP-2004-72, SWAT-}
\abstract{We study relevant deformations of an $\N=1$ superconformal
  theory which is an exactly marginal deformation of $U(N)$ $\N=4$ SUSY
  Yang-Mills. The resulting theory has a classical Higgs branch that
  is a  complex deformation of the orbifold $\C^3/\Z_n\times\Z_n$
  that is a non-compact Calabi-Yau space with
  isolated conifold singularities. At these singular points in moduli
  space the theory exhibits a duality cascade and flows to a confining
  theory with a mass gap. By exactly solving the corresponding
  holomorphic matrix model we compute the exact quantum
  superpotential generated at the end of the duality cascade and
  calculate precisely
  how quantum effects deform the classical moduli space by replacing
  the conifold singularities with three-cycles of finite size. Locally
  the structure is that of the deformed conifold, but the global
  geometry is different. This desingularized
  quantum deformed geometry is the moduli space of probe D3-branes at
  the end of a duality cascade realized on the worldvolume of
  (fractional) D3-branes placed at the isolated conifold singularities
  in the deformation of the orbifold $\C^3/\Z_n\times\Z_n$ with
  discrete torsion.}

\begin{document}

\section{Introduction}

Gauge theories with $\N=1$ supersymmetry (SUSY) in four dimensions are
known to exhibit rich and beautiful dynamics. The phenomenon of
Seiberg duality plays a central role in much of this
\cite{seiberg}. One of the most striking examples of the rich physics
of $\N=1$ SUSY theories is provided by the duality cascade phenomenon
uncovered by Klebanov and Strassler \cite{ks}. They investigated an
$\N=1$ gauge theory with $SU(M+N)\times SU(N)$ gauge group and
matter fields in the bi-fundamental representation which, upon
renormalization group (RG) flow, undergoes a succession of Seiberg
dualities or a duality cascade. The number of colors in each factor
drops repeatedly by $M$
units until finally in the infrared (IR) we are left with a pure
$\N=1$ SUSY Yang-Mills theory that confines and produces a mass
gap.\footnote{See the recent work
\cite{Gubser:2004qj} for further discussions on the IR
  universality class of these theories.}  In the ultraviolet
  (UV), however, these theories are not asymptotically free and hence are
not
  well-defined quantum field theories at all energy scales.

In this article, we study the phenomenon of a duality cascade within
a four dimensional UV-finite $\N=1$ SUSY field theory. Our study was
motivated by the
program initiated by Berenstein in \cite{ber, ber1, ber2}. The theory in
question is obtained from $\N=4$ SUSY Yang-Mills
theory with $U(N)$ gauge group by deformations involving exactly
marginal and relevant operators. Thus a UV-completion of the
Klebanov-Strassler cascade can be achieved within a
well-defined quantum field theory with gauge group of finite
rank. Indeed, the theory we study has discrete branches where it flows,
after a duality cascade, to an IR $\N=1$ gauge theory with a mass
gap. This IR theory (more precisely, its holomorphic sector) is
identical to the theory studied in \cite{mm3} which is quite similar
to $\N=1^*$ theory \cite{nickprem, ADK}. In the D-brane language, the
theories that we study are
realized on the worldvolume of D3-branes placed at isolated conifold
singularities of a Calabi-Yau threefold which, in this case, is a
complex deformation of the orbifold $\C^3/\Z_n\times\Z_n$. The
endpoint of the duality
cascade in the field theory has a geometric dual description where the
D-branes have disappeared and the Calabi-Yau geometry is deformed and
desingularized by the appearance of three-cycles of finite size. One
of our aims is to obtain the precise algebraic description of this quantum
deformed geometry with deformation parameters determined exactly in
terms of the parameters of the gauge theory. Although near the
singularities its local structure is that of the deformed conifold,
the global structure is very different. From this deformed
geometry we will also understand how the theory of \cite{ks} emerges
as a limiting case when one of the finite three-cycles in the geometry
is sent to infinity and we are left with a deformed conifold geometry.
The main tool that we will
employ to understand this deformed geometry (which is also the moduli space
of
probe D-branes at the end of the cascade)
and the gauge theory superpotential, is the holomorphic matrix
model approach of Dijkgraaf and Vafa \cite{DV1,  DV2, DV3}.

The theories of interest to us are obtained from the so-called
``$\beta$-deformation'' of the $\N=4$ theory with $U(N)$ gauge group. The
$\beta$-deformation constitutes one
of the two $\N=1$ supersymmetric, exactly marginal deformations of the
$\N=4$ RG fixed point identified by Leigh and Strassler in
\cite{LS}. It leads to
an $\N=1$ superconformal field theory (SCFT) with three adjoint chiral
supermultiplets $X$, $Y$ and $Z$ which obtain the tree level
superpotential interaction\footnote{We work with a normalization where the
kinetic terms of the chiral multiplets do not have a factor of ${1/
g^2_{YM}}$ in front.}
\EQ{
W=\lambda\Tr\,[XYZ - q XZY];\quad q:=e^{i\beta}\ .\label{superpot}}
where $\lambda$ and $q$ are complex bare parameters. We also introduce
the complex bare gauge coupling of this theory $\tau\equiv 4\pi
i/g^2_{YM}+\theta/2\pi$.
We will always restrict attention to the
case where $q$ is a primitive $n$-th root of unity, $q=\exp(2\pi
i/n)$. In order for the theory to be conformally invariant, the other
coupling $\lambda$ has to be some fixed (but unknown)
function of $\beta$ and the gauge coupling \cite{LS} such that as we
approach the $\N=4$ fixed line ($\beta\to0$), we also have
$\lambda\rightarrow 1$.
The gauge-coupling $\tau$ is a free
parameter of the theory and it was shown in \cite{mm3} that the
theory exhibits
electric-magnetic duality which acts as $SL(2,\Z)$
transformations on the gauge coupling:\footnote{Actually, in order to
  write the transformations in this way the gauge coupling undergoes
  an algebraic renormalization: we refer to \cite{mm3} for details.}
\EQ{
\htau\to\frac{a\htau+b}{c\tau+d}\
,\quad\beta\to\frac{\beta}{c\htau+d}\ .
}

Dynamical phenomena of interest ensue when we break conformal invariance by
perturbing the above $\N=1$ SCFT, with
$q=\exp(2\pi i/n)$, by certain relevant operators.
Specifically, we will consider adding to the tree
level superpotential, a mass term for one of the fields of the following
form
\EQ{W=\lambda\Tr\,[XYZ - e^{i{2\pi \over n}}
    XZY-\epsilon(\tfrac12X^2-aX)];\quad n>2\ .\label{deformation}}
More complicated polynomial deformations can also be considered
leading to similar IR dynamics, but these spoil the UV-behaviour. We
also consider slightly more general renormalizable deformation in the
penultimate section.

Let us now explain the essential ideas underlying the dynamics of the above
theories. It is well known that when $q$ is an $n$-th root of unity
the $\beta$-deformation of the $U(N)$, $\N=4$ theory can be
realized in IIB string theory as the worldvolume theory of
$N$ D3-branes placed at a $\C^3/\Z_n\times\Z_n$ orbifold singularity with
a single unit of discrete torsion \cite{Douglas1,Douglas2}.
The moduli space of vacua of the $\beta$-deformed theory contains
Coulomb and Higgs (and mixed) branches. It follows from the F-term
equations that, depending on the rank $N$ of the gauge
group, the Higgs branches within the moduli space are
symmetric products of a certain number of copies of
$\C^3/\Z_n\times\Z_n$.
(For example, when $N= mn$ there exists a Higgs
branch where the gauge group is broken to a $U(m)$ subgroup and the
Higgs branch is a symmetric product of $m$ copies of
$\C^3/\Z_n\times\Z_n$). The orbifold has three singular (complex)
lines which are fixed under the action of the orbifold group, and along
which the Coulomb and Higgs branches of the D-brane gauge theory
meet. The separations of the D-branes along the fixed lines
are the Coulomb branch moduli. At these singular fixed lines
there are fractional D-branes bound to the singularities\footnote{See
  \cite{Douglas2} for a discussion on the interpretation of these
  fractional branes.},
which can then annihilate in groups of $n$ to form a whole D-brane
free to
move off into the orbifold bulk (the Higgs branch.)

Once the theory has been deformed by relevant operators,
the fixed line singularities above can be
smoothed out by complex structure deformations. From the point of view of
the
gauge theory on the D-branes this corresponds to deforming the
superpotential. Deformations that resolve the fixed lines leaving
behind only isolated conifold singularities were first discussed in
\cite{Douglas2}. The deformation \eqref{deformation} that we consider
was discussed in \cite{ber2} and \cite{LB, LB1}. After a suitable
field redefinition it is equivalent to introducing masses for all
three multiplets, a deformation that has been studied for general
complex $\beta$ in \cite{mm3}. Importantly, for
the theories with $\beta=2\pi/n$, this class of deformations
does not lift the Higgs branch moduli space, rather it
leads to a moduli space which is a complex deformation of the
$\C^3/\Z_n\times\Z_n$
orbifold. Classically, the effect of the deformation is to smooth out two
of the three fixed lines into a single (complex) hyperboloid, while
the third fixed line is completely resolved leaving behind $[n/2]$
isolated conifold singularities.\footnote{Note that $[x]$ is the
  largest integer less than or equal to $x$.}
>From the string theory point of view,
once we have isolated conifold singularities in the spacetime geometry
we can imagine having
fractional branes stuck at these points. Indeed, in the field theory
perspective, a bulk D-brane is characterized by an $n$-dimensional
irreducible representation of the algebra following from the F-term
conditions implied by \eqref{deformation}. These representations
become reducible precisely at each of the
$[n/2]$ points in moduli space corresponding to the conifold
singularities. At any one of the conifold singularities labelled
by an integer
$p$, a bulk D-brane splits up into two lower dimensional
irreducible representations of dimension $p$ and $n-p$ respectively. These
are the two types of fractional D-branes bound to the singularity.

The appearance of isolated conifold singularities and fractional
branes pinned at them naturally points towards the
possibility of realizing $\N=1$ gauge theory dynamics of the kind
studied by Klebanov and Strassler
\cite{ks}. This was already realized by Berenstein in \cite{ber}
and \cite{ber2}. In this paper we will be able to make this
precise. First we will show that for the $U(N)$ gauge theory with
classical superpotential \eqref{deformation}, the low energy effective
theory at the isolated singular points is a $U(N^+)\times U(N^-)$
gauge theory (here $N=pN^+ + (n-p) N^-$) with bi-fundamental matter and
a quartic superpotential. In the D-brane language $N^\pm$ is the
number of fractional branes of each type.
In an appropriate regime of parameter space
which results in a separation of scales, this theory exhibits a duality
cascade that terminates in an $\N=1$ gauge theory with a mass gap. It
is important to note that the low-energy theory is not pure $\N=1$ SUSY
Yang-Mills theory. In fact it has additional massive fields and
is an $\N=1^*$ type theory which was studied in \cite{mm3}. In
particular the low energy theory develops a
superpotential and gaugino condensate both of which are non-trivial
modular functions of the microscopic gauge coupling $\tau$, the same
functions that were encountered in \cite{mm3}. Only in a
specific decoupling limit do we obtain pure $\N=1$ SUSY Yang-Mills.
All this is encoded rather beautifully in the geometry of the {\em
quantum deformed moduli space\/} which we are able to obtain exactly.

The classical moduli space has conifold singularities where
fractional branes are pinned. From general considerations
\cite{ks,mn,civ} in the quantum theory we expect the fractional branes
to deform the
geometry further. At the endpoint of the cascade we expect
that the (fractional) D-branes have disappeared, the conifold
singularities have
been replaced by $S^3$'s of finite size resulting in a quantum
deformed moduli space seen by a probe D-branes at the endpoint of
the cascade. Since all the holomorphic data of an $\N=1$ gauge theory
is encoded in a corresponding matrix model \cite{DV1, DV2, DV3}, we
should be able to determine this quantum deformed moduli space using the
matrix model approach. We achieve this using the fact that
the planar sector of the matrix model for our theory is in fact
exactly solvable \cite{mm3,mm1,mm2,kostov}. Remarkably, in the matrix
model one can understand the duality cascade of the field theory in
rather simple terms. It turns out that the matrix model description of
the field theory vacuum of interest with $U(N^+)\times U(N^-)$ gauge
symmetry, only depends on a single parameter corresponding to the glueball
superfield of the $U(|N^+-N^-|)$ gauge group factor appearing at the
end of the cascade. Here $|N^+-N^-|$ has the interpretation as the
total number of fractional branes pinned at the conifold singularity.
This particular feature was also encountered in
\cite{dvdecon} in the context of the matrix model description of $\hat A_1$
quiver theory of \cite{ks}.

For us the deformed Calabi-Yau geometry emerges
as a fibration over a Riemann surface that is completely determined by
the matrix model. This quantum geometry is a deformation of
the original classical Calabi-Yau space with conifold
singularities, and we find that the quantum deformation parameters can
be determined exactly as a function of the microscopic gauge coupling
$\tau$. The conifold singularities are now replaced by
three-cycles whose sizes are exactly calculable, nontrivial functions
of the gauge coupling or equivalently the glueball superfield.
An important feature of these deformed
geometries is that each of the ``blown-up'' $S^3$'s also has a finite
dual three-cycle. It is precisely in the limit that this dual cycle is sent
to infinity that we are left with just the usual deformed conifold
geometry which is dual to pure $\N=1$ SUSY Yang-Mills theory

The layout of this paper is as follows.  In Section 2, we review the
classical vacuum structure of the field theory and its classical
moduli space which has isolated singularities. We describe the
classical solutions that correspond to different types of fractional
branes stuck at singularities. In Section 3, we obtain the
classical effective action for the light modes at the singular points
in moduli space and observe that it corresponds (at weak coupling) to the
starting point of a duality cascade. In Section 4, by solving the
Dijkgraaf-Vafa matrix model associated to this theory we find the
quantum deformed moduli space where the deformation parameters are
determined precisely in terms of parameters of the gauge
theory. Section 5, considers theories with a slightly more general
renormalizable deformation including the terms $\Tr\,XY$ and
$\Tr\,XZ$. These theories can easily be solved along the lines we set
out in earlier sections.
Finally, in Section 6 we conclude with some remarks on
directions requiring further study. In Appendix A, we derive the
quadratic loop equations for the matrix models in question and point
out a way to obtain the general loop equations. Various useful
results involving elliptic functions are listed in Appendix B.


\section{Classical Vacuum Structure}

In this section we address the question of the classical vacuum structure of
the $\beta$-deformed theory with superpotential
\eqref{deformation}. In fact, following \cite{ber2} we will consider a
generalized version of the deformation in \eqref{deformation}.
Since we are primarily interested in the holomorphic
structure we shall follow the rule of solving the F-flatness
conditions modulo complex gauge transformations. Subject to certain 
well-known
caveats, such solutions are then equivalent to the solution of both
the F- and D-flatness conditions up to a complex gauge transformation.
%
%
%
We consider the gauge theory perturbed by a polynomial
superpotential\footnote{In the rest of the paper we set
  $\lambda=1$. It could easily be reinstated by by a re-scaling of the
fields.}
\EQ{
W= \Tr[XYZ - q XZY+V(X)]\ ,\quad
V(x)=\sum_{j\neq 0\text{ mod }n}^\ell {a_j\over j} x^{j}\ .
\label{pert}
}
The fact that we exclude terms in the superpotential
proportional to $X^j$, $j=0\text{ mod }n$ is the restriction of
\cite{ber2} and ensures that the deformation does not lift the Higgs
branch. Note that only
superpotentials for which $\ell=2$ are actually renormalizable and we
shall mostly have this case in mind because we want theories which
flow to the superconformal fixed-point in the UV. Note also that in this
case once we remove the linear term in $X$ by a shift, the
superpotential is equivalent to
\EQ{
W= \Tr\Big[XYZ - q XZY-\frac{a_1(1-q)}{a_2}YZ+\frac{a_2}2X^2]\ ,
}
which with a trivial
rescaling is precisely the theory investigated in \cite{mm3}.
However, for illustrative
purposes it is also useful to consider the $q=-1$ theory with $\ell=3$
where the superpotential is actually marginally irrelevant, or
non-renormalizable.

These perturbed theories and their moduli spaces of vacua have
been studied in \cite{ber1,ber2,LB,LB1}. In what follows, we shall
draw heavily on this analysis which shows that the classical moduli space of
vacua on a Higgs branch
of these theories corresponds to be a 3-complex dimensional Calabi-Yau
space $\X$ which is a complex deformation of the
orbifold with discrete torsion ({\it i.e.} a deformation of
$\C^3/\Z_n\times\Z_n$). We will further see that
the low-energy
description of the gauge theory involves both discrete vacua and a
moduli space with several branches. On some of these branches,
the gauge theory exhibits a subsequent
renormalization group cascade of the Klebanov-Strassler type
\cite{ks}, and on others the theory flows to the superconformal
fixed point of Klebanov and Witten \cite{KW}. In addition there are
isolated vacua where the theory confines without undergoing a duality
cascade. It is worth mentioning that the existence of this rich vacuum
structure is then inherited in the theories related by the $SL(2,\Z)$
group. In particular under S-duality, the Higgs branches are mapped to
confining branches of the dual theory with modified
gauge coupling $1/\htau$ and a deformation parameter $2\pi/(n\htau)$
\cite{mm3}.

\subsection{The case $q=-1$ with cubic deformation}

It is instructive to first specialize to the simplest example, namely the
theory
with $n=2$ (and $q=-1$) and subsequently describe the situation for $q$ a
generic $n$-th root of unity.
Consider then the superconformal theory with $q=-1$ in
Eq.~\eqref{superpot} perturbed by a tree
level superpotential for one of the fields $X$. We restrict attention
to a perturbation of the form
\EQ{V(x) = -\epsilon(x^3/3 -a^2 x).\label{restricted}}
This perturbation satisfies the restriction implied in
Eq.~\eqref{pert} which only requires that it should be a polynomial
in odd powers of $X$. As we discussed earlier,
a cubic superpotential would be marginally irrelevant once quantum
effects are taken into account. This fact renders this example less
interesting since it spoils the UV behaviour. However, it is useful to
discuss it because of its
simplicity and because it exhibits all the interesting infrared
dynamics that we are after. Later we shall consider theories
with $n>2$ with quadratic
superpotentials which exhibit the same properties but which are also
renormalizable. We shall also consider a more general class of
theories, including $q=-1$, which are renormalizable.

The F-term equations for the theory with $q=-1$ are
\EQ{
\{X,Y\}= \{X, Z\}=0\ ,\quad
\{Y,Z\}=\epsilon (X^2-a^2)\ .\label{fterma}
}
The classical moduli space of this theory can now be obtained
following the methods discussed
in \cite{LB, ber1, ber2}. First of all, complex gauge transformations
can be used to diagonalize, say, $X$. This leaves unfixed diagonal
gauge transformations as well as those generated by the Weyl group
which act by permutation of the two diagonal elements. We will have more
to say about these transformations shortly.
For the above choice of potential, the F-term equations are solved by
the following set of fields
\EQ{X= x\;\sigma^3;\;\;\;Y=y\;\sigma^2;\;\;\;Z = z\;\sigma^1 +
z^\prime\;\sigma^2\label{solutions}}
where the $\sigma^i$ are Pauli matrices with
\EQ{2\;y z^\prime = \epsilon\;(x^2-a^2).}
However, this solution is fixed by a $\Z_2\times\Z_2$ subgroup of the
gauge group (the product of the Weyl group and a subgroup of the
diagonal transformations). Explicitly the two generators are
\EQ{
\theta_1=\sigma_3\ ,\qquad\theta_2=\sigma_1\ .
}
Modding out by these transformations is equivalent to identifying
\EQ{
\theta_1:\quad (x,y,z)\sim(x,-y,-z)\ ,\qquad\theta_2:\quad
(x,y,z)\sim(-x,-y,z)\ .
\label{idents}
}

In the D-brane interpretation of \cite{LB, LB1}
this irreducible $2\times 2$ dimensional representation should be thought
of as parameterizing the moduli space of a single D3-brane probe moving
on a 3-complex dimensional Calabi-Yau space $\X$. In order to find
an explicit form for this classical moduli space we need to introduce
gauge invariant variables which automatically incorporate the
identifications \eqref{idents}.
This can be done
by first enumerating the set of elements that commute with all
the generators of the algebra in Eq.~\eqref{fterma}.
One finds $X^2$, $Y^2$ and $Z^2$. These 3 generators
are therefore proportional to the identity matrix and we can introduce
the gauge invariant variables $(u,v,w)$ via
\EQ{u\cdot 1_{\sst[2]\times[2]}
=X^2\ ,\quad v\cdot 1_{\sst[2]\times[2]}=Y^2\ ,\quad w\cdot
1_{\sst[2]\times[2]} =Z^2\ .}
Clearly
\EQ{
u=x^2\ ,\quad v=y^2\ ,\quad w=z^2+z^{\prime2}\ .}
There is an additional central element:
\EQ{t\cdot 1_{\sst[2]\times[2]}=XYZ + \tfrac12 X \;V^\prime(X)\ .}
The four variables $(t,u,v,w)$ are not independent, rather they
satisfy the relation
\EQ{t^2 = - uvw + {\epsilon^2\over 4}u(u-a^2)^2.\label{CYdef}}
This algebraic equation describes
a (non-compact) 3-complex dimensional Calabi-Yau space $\X$
in $\C^4$ which is precisely a
complex deformation of the
$\C^3/\Z_2 \times \Z_2$, the
orbifold with discrete torsion. Indeed, when the deformation
vanishes we have $t^2=-uvw$ which is the well-known algebraic
description of the orbifold.
The undeformed orbifold with discrete
torsion has (classical) singularities corresponding to the
(complex) fixed lines where any two of $u$, $v$ and $w$ vanish. These
fixed lines correspond to solutions \eqref{solutions} (with $\epsilon=0$)
fixed by $\theta_1$,
$\theta_2$ or $\theta_1\theta_2$, so $y=z=0$, $x=y=0$ and $x=z=0$,
respectively.

The addition of the superpotential perturbation ($\epsilon\neq 0$)
resolves the orbifold singularities, but only partially. In
particular, it is easy to see that the singular lines associated to the
{\it undeformed} solution fixed by the
gauge transformations $\theta_2$ and $\theta_1\theta_2$ ($u=v=0$ and
$u=w=0$, respectively) merge. For instance, the
singular line associated to $\theta_1\theta_2$ becomes $x=z=0$ and
$yz'=-\epsilon a^2/2$, or $u=0$ and $wv=(-\epsilon a^2/2)^2$.
However, the third singular
line associated to the solution fixed by $\theta_1$ is reduced to
an isolated
singular point located at $(u=a^2,v=0,w=0)$. This singularity is locally of
  the conifold type.\footnote{This can be seen by introducing the new
variable
  $\delta = u-a^2$ and focusing attention on $\delta,v,w\ll a^2$.} In fact
  precisely as we approach this singular point from the bulk, the
  two-dimensional solutions to the F-term
  equations above become reducible in the sense that there are now two
  one-dimensional solutions $y=z=0$ and $x=\pm a$. In the brane
  interpretation, the fact that the
  two-dimensional irreducible representations of the deformed algebra
  split up into two one-dimensional representations can be interpreted
  as the fractionation of a bulk brane at the conifold singularity. In
  particular each one-dimensional solution corresponds to a fractional
  brane stuck at the conifold singularity while the irreducible
  two-dimensional solution describes the possibility for the
two types of fractional branes
  at the singularity to combine into a whole
  brane and move away from the singular point into the bulk of the
  deformed geometry which is also the Higgs branch of the D-brane gauge
theory. (Note that the one-dimensional solutions only exist at $u=a^2$
and so an individual fractional brane can only be pinned at this
singularity.) In the superconformal theory with $\epsilon=0$ this
conifold singularity becomes a point on the fixed line $v=w=0$ where
the Coulomb branch meets the Higgs branch.

\subsection{The general case $n>2$}

The picture that emerges for generic $n>2$ is quite similar. If we are
only interested in the low-energy dynamics we can always allow arbitrary
polynomial perturbations satisfying the requirement \eqref{pert}
resulting in classically deformed moduli spaces that will be described
below. The
restriction to renormalizable terms will become important in the
context of what we have to say in the next section.

The F-term equations for the perturbed theory are
\EQ{
XY-qYX=0\ ,\quad ZX-qXZ=0\ ,\quad YZ-qZY=-V'(X)
\label{fterm}}
To start with there are two Coulomb branches where $X=Z=0$ and $Y$
diagonal and $X=Y=0$ and $Z$ diagonal, respectively. We will not be
interested in these branches. The remaining solutions can be built out
of irreducible blocks of size up to $n$. Blocks of dimension $p<n$
come in sets $\{\RR^{(p)}_i\}$ while the block $\RR^{(n)}$
of dimension $n$ is unique. In the $q=-1$ case above these were the two
one-dimensional solutions and the two-dimensional irreducible
solution, respectively. The solutions
with dimension $p<n$, the $\{\RR^{(p)}_i\}$, will describe
fractional branes pinned at conifold singularities. The
other irreducible $n$-dimensional solution $\RR^{(n)}$ will describe
whole branes that are free to move off the
singularities and explore the bulk of the classical moduli space.

In the following a central r\^ole will be played by the function $f(x)$
defined in terms of the potential by
\EQ{
xV'(x)=f(x)-f(xq)\ .
\label{deff}
}
Note that
\EQ{
f(x)=\sum_{j\neq
  0\text{ mod }n}^\ell\frac{a_jx^j}{1-q^j}}
and has the same order as $V(x)$. The
two classes of solutions to \eqref{fterm} are more specifically

(i) $\RR_i^{(p)}$, $i=1,\ldots,d^{(p)}$,
of dimension $p<n$:
\EQ{
X^{(p)}_i=e^{(p)}_i\MAT{1&&&&\\ &q&&&\\ &&q^2&&\\  &&&\ddots&\\
&&&&q^{p-1}},\
Y^{(p)}_i=\MAT{0 &&&& \\ 1&0& && \\ & 1& 0 & &\\
&&\ddots & \ddots & \\ &&& 1&0},\
Z^{(p)}_i=\MAT{0&z_1 &&& \\ &0&z_2 && \\ &&\ddots &\ddots &\\
&&& 0 & z_{p-1}\\ &&&&0},
\label{reduced}}
where $e^{(p)}_i$ is a root of the polynomial equation
\EQ{
f(x)=f(xq^p)
}
excluding $x=0$.

(ii) $\RR^{(n)}[u=x^n,v=y^n,w=z^n]$, of dimension $n$:
\EQ{
X^{(n)}=x\MAT{1&&&&\\ &q&&&\\ &&q&&\\ &&&\ddots&\\ &&&&q^{n-1}},\
Y^{(n)}=y\MAT{0 &&&&1 \\ 1&0& && \\ & 1& 0 & &\\
&&\ddots & \ddots & \\ &&& 1 &0},\
Z^{(n)}=z\MAT{0&\zeta_1 &&& \\ &0&\zeta_2 && \\ &&\ddots &\ddots &\\
&&& 0 & \zeta_{n-1}\\ \zeta_n&&&&0}
}
with $\prod_{j=1}^n\zeta_j=1$. The actual values of the $\zeta_i$ are
fixed but will not be required.

Since $x$, $y$ and $z$ are not fixed
the representation $\RR^{(n)}[u,v,w]$ has a moduli space which
can be obtained, as in the $q=-1$ case,
by determining the relation between the elements of the
center of the algebra, namely
\EQ{u\cdot 1_{\sst[n]\times[n]}= X^n\ ,\quad
v\cdot 1_{\sst[n]\times[n]}=Y^n\ ,\quad
w\cdot 1_{\sst[n]\times[n]}=Z^n\ ,\quad t\cdot
1_{\sst[n]\times[n]}=XYZ+f(X)\ .}
The set of variables $(t, u, v, w)$ satisfy the relation
\EQ{
uvw=(-1)^{n+1}\prod_{j=1}^n\big(t-f(xq^j)\big)=
(-1)^{n+1}F(t,u)\ .\label{deformed}
}
Notice that the right hand side of this equation is actually invariant under
$x\rightarrow q x$ and can therefore be only a polynomial $F(t,u)$ in the
invariant $u=x^n$. This is a 3-complex dimensional Calabi-Yau space
$\X$ which is a deformation of the orbifold
$\C^3/\Z_n\times\Z_n$ with discrete torsion described by the algebraic
equation $t^n=(-1)^{n+1}uvw$.

As in the $q=-1$ case the singular fixed lines of the orbifold are
partially resolved. Rather than investigating these singularities by
considering the action of gauge
transformation as we did for $q=-1$, the structure of the
singularities can equivalently be
determined directly from the Calabi-Yau geometry \eqref{deformed},
as in \cite{ber2}, by demanding that the partial derivatives with
respect to $u$, $v$, $w$ and $t$ vanish. The first kind correspond to
$u=0$ along with
\EQ{
vw=\partial_uF(t,u)\big|_{u=0}\ ,
}
which is a deformation of the two fixed lines $u=v=0$ and
$u=w=0$. The other class of singularities are the ones that interest
us. They are isolated points in the geometry where
$v=w=0$ and the the $n$-sheeted algebraic curve, or Riemann surface,
$\Sigma$ defined by
\EQ{
F(t,u)=0
\label{defs}
}
becomes singular.
These points are simply given by the roots of
\EQ{
f(x)=f(xq^p)
}
for any $p < n$, excluding $x=0$.
Hence the isolated singular points are precisely at
the locations of the
fractional branes associated to the representation $\RR^{(p)}_i$. It

Earlier
we denoted these roots
$e^{(p)}_i$, where $i$ counts the degeneracy:
$i=1,\ldots,d^{(p)}$. The fractional branes are
pinned at a conifold-type singularity at
\EQ{
u=u^{(p)}_i\equiv\big(e_i^{(p)}\big)^n\ ,\quad v=w=0\ .
}
Let us enumerate the number of these isolated singular
points. Recall that we
are counting the roots of
$f(x)-f(xq^p)$ which is a polynomial equation of degree $\ell$, except when
$\ell
p=0\text{ mod }n$, in which case it has degree $\ell-1$. So the number
of roots, apart from $x=0$, is
\EQ{
d^{(p)}=\begin{cases}\ell-2 & \ell p=0\text{ mod }n\\ \ell-1
&\text{otherwise}\ .\end{cases}
}
Note also that there is a symmetry
\EQ{
e^{(n-p)}_i=q^pe^{(p)}_{d^{(p)}+1-i}
}
and $d^{(p)}=d^{(n-p)}$. (These equivalences are valid
when $p=n/2$ for $n$ even.) This means that the fractional branes
associated to $\RR^{(p)}_i$ and $\RR^{(n-p)}_{d^{(p)}+1-i}$ lie at the same
point in the geometry:
\EQ{
u^{(p)}_i=u^{(n-p)}_{d^{(p)}+1-i}\ .
}
This is very familiar from the brane interpretation of the
basic conifold theory: fractional branes come in pairs.
In order to avoid this cumbersome notation, we shall label the
singularities, or pairs of mirror fractional branes, by the index
$\mu=(p,i)$. In particular, for a given $\mu$ the associated value of
$p$ is $1\leq p_\mu\leq [n/2]$ and we denote the mirror pair
$\RR^{(p)}_i$ and $\RR^{(n-p)}_{d^{(p)}+1-i}$ as $\RR^+_\mu$ and
$\RR^-_\mu$. Finally
\EQ{
u_\mu\equiv u^{(p)}_i=u^{(n-p)}_{d^{(p)}+1-i}\ ,\qquad\mu=1,\ldots,g\ .
}
Note that the number of singularities, or pairs of fractional branes,
is equal to
\EQ{
g=\frac{d^{(n/2)}}2+\sum_{j=1}^{[(n-1)/2]}d^{(p)}
\label{defg}
}

However, can we see that if we have two fractional branes,
one of each type, then it can become a ``complete'' brane and move off
into the bulk? The answer is yes because at $(u=u_\mu,v=0,w=0)$
the representation $\RR^{(n)}[u,v,w]$ is actually reducible:
\EQ{
\RR^{(n)}[u=u_\mu,v=0,w=0]\simeq \RR^+_\mu\oplus
\RR^-_\mu\ .
}

Let us consider in more detail the relation of these singularities to
the algebraic curve $\Sigma$ defined by $F(t,u)=0$. We can think of
$\Sigma$ as an $n$-fold cover the $u$ plane. There is a
$n^\text{th}$-order branch cut
running from $u=0$ to $u=\infty$ which joins all the sheets. In
addition for $u=u_\mu$ a pair of sheets
touch at a point. So as one expects the curve $\Sigma$ is singular at
these points. In the quantum theory the picture emerging will be that the
presence of a net number
fractional branes, {\it i.e.\/}~the number of representations
$\RR^+_\mu$ minus the number of $\RR^-_\mu$,
has the effect of deforming the isolated conifold singularity in
$\X$ which at the level of the curve $\Sigma$
involves the opening up of the point $u=u_\mu$
into a cut joining the pair of sheets.

Now that we have described the irreducible blocks, we can now write
down a general classical vacuum. If one starts with a $U(N)$ gauge
group then a general vacuum is associated to a reducible representation
\EQ{
\oplus_{\mu=1}^g\Big(N^+_\mu\cdot \RR_\mu\oplus N^-_\mu\cdot
\RR^-_\mu\Big)\ ,
\label{vaca}
}
where
\EQ{
N=\sum_{\mu=1}^g\Big(p_\mu N^+_\mu+(n-p_\mu)N^-_\mu\big)\ .
}
Notice that we didn't include the $\RR^{(n)}$ block explicitly because
the vacuum \eqref{vaca} lies at a point in a large moduli space where
up to $|N^+_\mu-N^-_\mu|$, for each pair $\RR^\pm_\mu$
of fractional branes, can move of into the bulk and explore the space $\X$.

\section{The Classical Effective Action and RG Cascade}

We have seen that certain deformations of the
$\beta$-deformed $\N=4$ theory lead to deformed (classical) moduli
spaces for D-branes with isolated
conifold singularities where fractional branes can be stuck.
We will now ask what the effective gauge theory is on stacks of
branes some of which remain stuck at the isolated points while others
have moved off to explore the full moduli space. As in the last Section
it is useful to consider the $q=-1$ case first.

\subsection{The case $q=-1$ with cubic deformation}

Consider the
$U(N)$ gauge theory with the cubic plus linear superpotential
\eqref{restricted}.  Recall that there is a single isolated singularity
in $\X$ at $u=a^2$, $v=w=0$. The two fractional branes are
distinguished by $x=\pm a$. So let us consider a vacuum with
$N^+$ (fractional) branes with $x=+a$ and $N^-$ branes with
$x=-a$ (fractional branes of the other type) with $N^++N^-=N$. In this
vacuum
\EQ{X = \MAT{a\cdot1_{\sst [N^+]\times [N^+]}
& 0\\ 0 &
-a\cdot1_{\sst [N^-]\times[N^-]}}\ ,\qquad Y=Z=0\ .
\label{vacuum}
}
Note that as we remove the deformation smoothly
$V\rightarrow 0$, these points lie at the intersection of the Coulomb
and Higgs branches of the parent superconformal theory.
The classical vacuum \eqref{vacuum} preserves a $U(N^+)\times U(N^-)$
gauge symmetry.

The low-energy classical $U(N^+)\times U(N^-)$ gauge theory at these points
in moduli space is obtained by expanding in fluctuations about the
above solution,
\EQ{\delta X=\MAT{X_1
& 0\\ 0 &
X_2}, \quad \delta Y=\MAT{Y_1 & A_1\\ B_1 &
    Y_2},\quad \delta Z= \MAT{Z_1 & A_2\\ B_2 &
    Z_2}\ ,}
where $X_1$, $Y_1$ and $Z_1$ transform in the adjoint representation
of the unbroken $U(N^+)$, while $X_2$, $Y_2$ and $Z_2$ are in the
adjoint of the $U(N^-)$. In addition the fields $A_i$ and
$B_i$ fill out four bi-fundamental multiplets transforming as
$(N^+,\bar{N}^-)$ and $(\bar{N}^+, N^-)$ respectively.

The nature of the low-energy theory can now be easily determined by
expanding out the classical superpotential in terms of the fluctuations
and we find
\SP{W&= \Tr[X_1(A_1 B_2 + A_2 B_1)-\epsilon a
X_1^2-{\epsilon\over 3}X_1^3+
2a Y_1 Z_1+ X_1( Z_1 Y_1+
    Y_1 Z_1)]\\
&+ \Tr[X_2
(B_1 A_2 + B_2 A_1)+\epsilon a X_2^2-{\epsilon\over 3}X_2^3
-2 a Y_2 Z_2 + X_2(Y_2 Z_2 + Z_2 Y_2) ]\ .
\label{supeff}
}
Note that the fluctuations $Y_{1,2}$ and $Z_{1,2}$ have a mass set by
the scale $  a$ while those associated to $X_1$ and $X_2$ have a
mass set by the scale $ \epsilon a$. Assuming $\epsilon\ll1$ and
the gauge coupling is weak enough, it
makes sense we can integrate out $Y_{1,2}$ and $Z_{1,2}$ at the
classical level to arrive at the effective superpotential
\EQ{W_\text{eff}= \Tr[X_1(A_1 B_2 + A_2 B_1)-\epsilon
    aX_1^2]+ \Tr[X_2
(B_1 A_2 + B_2 A_1)+\epsilon a X_2^2]\ ,
}
where we have neglected the cubic coupling of $X_{1,2}$ which will be
irrelevant in the IR. The effective theory for the light modes is a
mass deformation of the $U(N^+)\times U(N^-)$ $\N=2$ theory with equal
and opposite sign masses for the two adjoint scalars.
This is precisely the
starting point of the Klebanov-Strassler duality cascade.
We can also now integrate out $X_{1,2}$ to arrive at
\EQ{W_\text{eff} = -{ 1\over 4\epsilon a}\Tr(A_i B_j A_k
  B_l)\epsilon^{ik}\epsilon^{jl}\ .
}
For a fixed given value of $N$ we can vary
$N^+$ (or $N^-$) and explore different vacua of the theory. In particular
when the unbroken rank $N$ is even we can choose
the vacuum with $N^+=N^-=N/2$. Here the resulting $U(N/2)\times U(N/2)$
theory with
bi-fundamentals and the quartic superpotential above flows to the
superconformal conifold theory of Klebanov and Witten \cite{KW}. As
pointed out in \cite{KW},
the quartic superpotential is actually an exactly marginal operator at
the IR fixed point. Thus we have found a flow that interpolates between
the ${\C^3/\Z_2\times\Z_2}$ theory and the Klebanov-Witten conifold
theory.

For any given $N=N^++N^-$ we can sweep through distinct classical
vacua by varying $N^+$. In particular, when $N^+=0$, we simply have a $U(N)$
gauge theory with $\N=1$ SUSY with massive adjoint scalars. The
$SU(N)$ subgroup of the $U(N)$ gauge group
confines and generates a mass gap
and in many ways is similar to the confining vacuum of $\N=1^*$ gauge
theory \cite{nickprem, PS}. In an appropriate decoupling limit for the
adjoint scalars, the theory reduces to pure $\N=1$ SUSY gauge theory.

The remaining vacua labelled by generic values of $N^+$ contain
rich and complicated dynamics. In particular, at least at
weak coupling $g^2_{YM}\ll1$ (the gauge-coupling of the parent
$q=-1$ SCFT), the low-energy $U(N^+)\times U(N^-)$ theory
with bi-fundamentals and a quartic superpotential shares all the
features of the theory studied by Klebanov and Strassler
\cite{ks}. Under renormalization group flow, the individual gauge
group factors will undergo successive Seiberg dualities that
terminate at an $\N=1$ theory with a mass gap and gaugino
condensation, with or without probe branes
propagating on a deformed and desingularized moduli space. We will
return to a more detailed quantum description of
these vacua after demonstrating how they arise in the theories with
$n>2$.

\subsection{The case $n>2$ with quadratic deformation}

At the outset we remark that there is a key
difference between the cascade theories encountered here (and above)
and that of
Klebanov-Strassler. The theory considered in \cite{ks} has the
property that one of the gauge group factors has a Landau pole in the
ultraviolet. (One could imagine achieving a UV completion by
a reverse duality cascade, but in that case the ranks of the gauge
group factors
would grow without bound in the UV \footnote{For a discussion of the
  UV behaviour of various theories exhibiting the phenomenon of
  duality cascades, see\cite{ami, fiol}}.) The theories we consider have the
property that
they are obtained as deformations of a four dimensional
$\N=1$ superconformal field theory with $U(N)$ gauge group and
therefore any potential duality cascade is embedded within a
well-defined UV-theory. In this context, note that although
the deforming potential for $q=-1$ was classically marginal
({\it i.e.\/} cubic), it is most likely to be rendered irrelevant by quantum
effects. However, as we will see below for the theories with $n>2$ the
deforming potentials need only be quadratic---hence
renormalizable---to obtain an RG cascade in the infrared.

We consider the superconformal theory perturbed by the superpotential
\EQ{V(x)=-\epsilon(x^2/2 -a x).\label{mass}}
This superpotential yields exactly one fractional brane solution
for each irreducible representation of dimension $p<n$ (except for
$p=n/2$ when $n$ is even) stuck at
\EQ{
u=u^{(p)}=\Big(a \frac{1+q}{1+q^p}\Big)^n\ ,\quad
v=w=0\label{singularities}}
in $\X$.
We now consider the theory with a $U(N)$, $N=pN^++(n-p)N^-$ gauge group
in a vacuum with $N^+$ fractional
$\RR^{(p)}$ branes and $N^-$ fractional $\RR^{(n-p)}$
branes. We are going to argue that as in the $q=-1$ case
the classical effective theory is a $U(N^+)\times U(N^-)$ theory with
massless bi-fundamentals.
First-of-all, the gauge group is obviously broken to $U(N^+)\times
U(N^-)$. This is clear from the from of the VEV assigned to the adjoint
field $X$:
\EQ{
X = \MAT{1_{\sst[N^+]\times[N^+]}\otimes X^{(p)}
& 0\\ 0 &1_{\sst[N^-]\times [N^-]}\otimes X^{(n-p)}}\ .
\label{vac}}
But the question is whether there are the right number of massless
bi-fundamental fields. Indeed the answer is yes, as we can see by first
considering the simpler
case of the theory realized on one $\RR^{(p)}$ brane
and one $\RR^{(n-p)}$ brane. As we have already seen, these two
fractional branes make a complete brane that can move off onto the Higgs
branch and thus there must be additional massless moduli. These
additional moduli correspond to the non-zero elements of $Y$ and
$Z$ in the representation $\RR^{(n)}[u,v,w]$ that are zero in
$\RR^{(p)}\oplus \RR^{(n-p)}$.
More specifically if $E_{i,j}$ is a matrix with a
1 in position $(ij)$ and zeros elsewhere, then the elements in
question are $E_{p+1,p}$ and $E_{1,n}$ in $Y$ and $E_{p,p+1}$ and
$E_{n,1}$ in $Z$.

Therefore, when we have $N^+$ coincident $\RR^{(p)}$ branes
and $N^-$ $\RR^{(n-p)}$ branes at the singular point there will be
$4\times N^+\times N^-$ such massless
fields which will fill out a bi-fundamental multiplet
in the
$(N^+,\bar{N}^-)\oplus(\bar{N}^+,N^-)$ of the low-energy product
gauge group.
To be more specific the low-energy degrees of freedom can be encoded
in the following set of fluctuating modes
\AL{
\delta X&=\MAT{X_1\otimes P^{(p)} & 0\\ 0 & X_2\otimes
q^pP^{(n-p)}}+\tilde\delta X\ ,\\
\delta Y&=\MAT{0& A_1\otimes E_{1,n-p}
\\
B_1\otimes E_{1,p}&0}+\tilde\delta Y\ ,\\
\delta Z&=\MAT{0 & q^{-p}A_2\otimes E_{p,1}
\\
B_2\otimes E_{n-p,1}&0}+\tilde\delta Z\ .
}
Here  $X_1$ transforms in the adjoint representation of the
unbroken $U(N^+)$ while $X_2$ transforms in the  adjoint of
$U(N^-)$. We have also defined $P^{(p)}=\text{diag}(1,q,\ldots,q^{p-1})$.
The modes $(A_i, B_i)$ are the
bi-fundamentals transforming in the $(N^+,\bar{N}^-)$ and
$(\bar{N}^+,N^-)$ representations, respectively. The remaining
fluctuations (modulo gauge transformations) are denoted $\tilde\delta
X$, {\it etc.\/}.
The classical superpotential for the fluctuations at
the chosen vacuum is then
\SP{
\frac1  W_\text{eff}
&=\Tr\big(X_1(A_1B_2-A_2B_1)\big)+\Tr\big(X_2(B_1A_2-B_2A_1)\big)\\
&\qquad\qquad
+\frac\epsilon2\cdot\frac{1-q^{2p}}{1-q^2}\big(\Tr\,X_1^2-\Tr\,X_2^2\big)
+\tilde\delta W\ .\label{effw}
}
The remaining factor $\tilde\delta W$ includes all the additional
fluctuating modes.
Importantly all these modes gain a mass at a scale
$a $ and so for sufficiently large $a$ compared with $\epsilon$
there is a separation-of-scales
between these tilded modes and the ones we have singled out and we can
ignore them---at least at sufficiently weak coupling. Note that the
effective couplings of the $U(N^+)$ and $U(N^-)$ factors are $p\tau$
and $(n-p)\tau$, respectively.

We might have expected the above picture to be true since from a
geometric point-of-view taking $a\gg\epsilon$ zeros-in on the
singularity which is then well approximated by the deformed conifold
and the usual story of the duality cascade and deformed conifold
should then be applicable \cite{ks}. Indeed
upon integrating out the massive adjoints we obtain precisely, the
quartic superpotential
\EQ{W_\text{eff} = -\frac{1 }{2\epsilon}\cdot\frac{1-q^2}{1-q^{2p}}
\Tr(A_i B_j A_k B_l)\epsilon ^{ik}\epsilon^{jl}\ .}
Note that we have obtained what will be the starting point for a
duality cascade under renormalization group flow, specifically when the
gauge coupling of the ultraviolet theory is taken to be sufficiently
weak in order that the effective superpotential \eqref{effw} is valid
over a certain range of energies between the decoupling of the
additional modes and the onset of the cascade. What is new and
remarkable is that the duality
cascade has been embedded in a four-dimensional
theory with a perfectly well-defined UV limit given by the
superconformal fixed point. The RG flow away from the fixed point is
defined by the mass deformation \eqref{mass}. The end-point of the cascade
is a
$U(|N^+-N^-|)$ theory which confines. As noted earlier, when $N^+=N^-$
the theory flows to the conifold superconformal theory realizing a
flow between the $\C^3/\Z_n\times \Z_n$
orbifold and the Klebanov-Witten SCFT.

It is clearly possible to generalize the discussion to a vacuum with
more than one type of fractional brane pair. In this case, at low energy the
theories associated to each pair will decouple from each other.

\section{Engineering the Quantum Geometry from the Matrix Model}

We have seen how the classical low-energy description of the perturbed
SCFT
leads to a picture of the gauge theory being realized on
(fractional) D-branes placed at conifold singularities in a geometry $\X$
which is a deformation of the $\C^3/\Z_n\times
\Z_n$ orbifold. This deformed geometry was precisely the
Higgs branch moduli space of the gauge theory.
>From the classical picture we also concluded
that at generic classical vacua the theory undergoes
strong-coupling quantum dynamics such as successive Seiberg dualities
leading to a duality cascade and/or confinement with gaugino
condensation. We expect that the nonperturbative quantum effects
({\it i.e.\/}~the presence of fractional branes) should result in a further
deformation of the geometry above which desingularizes the isolated
conifold points. This was exactly the picture at the endpoint of
the RG cascade in \cite{ks} wherein the associated geometry was just
the deformed conifold. But this cannot be the whole story in the class
of theories that we are studying as they are actually
UV finite theories that confine in the IR with or
without an intermediate duality cascade. In a certain decoupling limit
for the adjoint scalars they will reduce precisely to the theory of
\cite{ks}. Correspondingly, the deformed
Calabi-Yau geometry $\X$
(which will be the moduli space seen by probe branes at
the end of a cascade) will only reduce to the deformed conifold in a limit
where some 3-cycle is sent to infinity.

>From the work of Dijkgraaf and Vafa \cite{DV1,DV2,DV3}, it is now
well-understood how to
obtain the geometry associated to a given ${\cal N}=1$ gauge theory. In
particular, one engineers the gauge theory on D-branes in
a Calabi-Yau geometry. Via a geometric transition the system has a
dual description in terms of a deformed Calabi-Yau geometry with
fluxes and no D-branes and this geometry is captured by a certain
Riemann surface encoded in a specific matrix model.

The matrix model associated to our theory is defined by the partition
function
\EQ{
{\cal Z}= {1\over {\rm Vol}[U(\hat N)]}\int [dX]_{{\hat N}^2}
  [dY]_{{\hat N}^2} [dZ]_{{\hat N}^2}\;
{\rm exp}[-\frac{1}{g_s}\Tr(XYZ-qXZY+V(X))]\ .
}
where there is matrix for each chiral field of the theory. As usual
the dimensions of the matrices indicated by
$\hat N$ are not the same as the fields in the parent field theory.

As in earlier works \cite{mm3,mm1,mm2} and references therein, the
holomorphic integral is to be thought of as a contour integral and in
practice is performed
by choosing $Y=Z^\dagger$ and then integrating them out.
The resulting one-matrix model can be
re-written as an ordinary $\hat
N$-dimensional integral in terms of the eigenvalues
$\{x_i\}$ of $X$:
\EQ{
{\cal Z}=\int \prod_{i=1}^{\hat N}
  dx_i\;
\frac{\prod_{i\neq j}x_i-x_j}{\prod_{i,j}x_i-qx_j}
\;e^{-{1\over g_s}\sum_iV(x_i)}\ .
\label{eigenintegral}
}

\subsection{Saddle point condition}

The connection with the holomorphic sector of the gauge theory and the
associated Riemann
surface emerges from the so-called genus zero contributions to the
matrix integral when evaluated in a 't Hooft large-$\hat N$ expansion
around a saddle-point. In this limit, $\hat
N\to\infty$ and $g_s\to0$ with $S=g_s\hat N$ fixed. The saddle
point configuration of the eigenvalues $\{x_i\}$ is determined
by extremizing the effective action \eqref{eigenintegral} leading
to the condition
\EQ{
g_s\Big[2\sum_{j\neq i} {1\over
{x_i-x_j}}
-\sum_{j} {1\over {x_i-qx_j}}-\sum_{j} {1\over {x_i-q^{-1}x_j}}\Big]=
  V^\prime(x_i)\ .\label{spe}
}
In practice,
one starts with a classical saddle point solution (of the field theory),
where the left-hand side of \eqref{spe} vanishes,
described by a set of
eigenvalues $\{x_a\}$ with a certain degeneracy $N_a$, $\sum_aN_a=N$.
In the matrix model, one replaces $N_a\to\hat N_a$, $\sum_a\hat
N_a=\hat N$, and takes $\hat N_a\to\infty$ and $g_s\to0$
with each $\hat S_a=g_s\hat N_a$
fixed. In this limit, the basic assumption is that the eigenvalues at
$x_a$ spread out in a one-dimensional distribution on an open contour
$\CC_a$ in the complex $x$-plane. This distribution can be described by
a unit-normalized density $\rho(x)$ with support on the union of contours
$\CC=\cup_a\CC_a$. The density of eigenvalues
can be encoded in the resolvent function
\EQ{
\omega(x)=\lim_{\hat N\rightarrow \infty}\frac1{\hat N}
\sum_i\frac1{x-x_i}=\int_{\CC} dy \frac{\rho(y)}{x-y}\ .
}
The resolvent $\omega(x)$ is
analytic and only has cuts along the equilibrium distribution with a
discontinuity that determines the density:
\EQ{
\omega(x+\epsilon)-\omega(x-\epsilon)=2\pi i\rho(x)\ ,\qquad x\in\CC\ ,
}
where $\epsilon$ is a suitable infinitesimal. Integrating the resolvent over
any one of its branch cuts $\CC_a$ gives us the fraction of
the total number of eigenvalues, $\hat N_a/\hat N$ , that are distributed
along that cut.

The saddle-point equation
\eqref{spe} in the large $\hat N$ limit can now be written in terms of
the resolvent function as
\EQ{
S\;[\omega(x+\epsilon)+\omega(x-\epsilon)-q\omega(qx)-q^{-1}\omega(q^{-1}x)]=
  V'(x)\ ,\qquad
x\in\CC\ ,
\label{nspe}
}
where $\epsilon$ is a suitable infinitesimal so that $x\pm\epsilon$
lie above and below the cut at $x\in\CC$.
The content of this saddle point equation is easily understood
when recast in terms of a new function $t(x)$ (see
also \cite{mm3,mm1,mm2,kostov}) which naturally
encodes the geometry of the quantum deformed moduli space and will also
allow us to see the duality cascade of the field theory in a simple way.
The form of the saddle point equation \eqref{nspe} motivates us to define
\EQ{
t(x):=f(x)+S\;x\;\big(q^{-1}\omega(q^{-1}x)-\omega(x)\big)\ ,
\label{deft}
}
where $f(x)$ was introduced in \eqref{deff}. From the analytic
structure of the resolvent $\omega(x)$ it follows that $t(x)$ has cuts at
$\CC=\cup_a\CC_a$ and its rotation by $q$: $q\;\CC=\cup_a(q\;\CC_a)$.
The saddle-point equation \eqref{nspe} is then very simple:
\EQ{
t(x\pm\epsilon)=t(q(x\mp\epsilon))\ ,\qquad x\in\CC\ .
\label{spet}
}
This is simply a gluing condition that glues a point $x\pm\epsilon$
on the top/bottom of the cut $\CC_a$ to a point
$q(x\mp\epsilon)$ on the bottom/top of the cut $q\;\CC_a$. So $t$ defines
a Riemann surface $\tilde\Sigma$ which is a copy of the complex
$x$-plane with the cuts identified as above.

Before we describe the solution of the saddle-point equations and the
associated Riemann surface $\tilde \Sigma$ it is useful to
develop an intuitive picture anticipating some of the key features of
the function $t(x)$. We first recall from the results of the previous
section that in a field theory vacuum
preserving a $U(N^+) \times U(N^-)$ gauge symmetry, the field $X$
has the classical eigenvalues $e_\mu (1,q,\ldots,q^{p-1})$ with a
degeneracy of
$N^+$ for each, and $e_\mu (q^p,q^{p+1},\ldots, q^{n-1})$ with a
degeneracy $N^-$. Correspondingly, in the classical limit $g_s\rightarrow 0$
the matrix model resolvent will have simple
poles at $x\in\{e_\mu,qe_\mu ,\ldots,q^{p-1}e_\mu\}$
with residue $\hat N^+/\hat
N$, and at
$x\in\{q^pe_\mu,q^{p+1}e_\mu,\ldots\, q^{n-1}e_\mu\}$
with residue $\hat N^-/\hat N$. This
means that in the classical limit the function $t(x)$ defined in
\eqref{deft} has only {\em two\/} poles, at $x= e_\mu$  and
$x=e_\mu q^p$ with residues $g_s(\hat N^+-\hat N^-)e_\mu$ and $g_s(\hat
N^+-\hat N^-) q^pe_\mu$
respectively. Thus, in this simplified classical limit, we can see
that the matrix model depends only on one ``modulus'' $g_s(\hat
N^+-\hat N^-)$.
The nontrivial result
which will be argued below is that when the matrix model
interactions are turned on ($g_s\neq 0$), each of these two poles will
get smeared
into a {\em single} branch cut of $t(x)$ and the saddle point condition
glues these two cuts together. Importantly, the resulting Riemann
surface $\tilde\Sigma$ only depends on the modulus $g_s(\hat N^+-\hat
N^-)$. This
modulus is naturally interpreted as the gluino condensate in the
$U(|N^+-N^-|)$ gauge theory on fractional branes at the end of
the duality cascade in the field theory. A similar matrix model
interpretation was found in \cite{dvdecon} for
the duality cascade of the $\hat A_1$ quiver theory.

\subsection{The general solution}

Let us now turn to the solution of the saddle point equation which
will determine the Riemann surface $\tilde\Sigma$. We will keep the
discussion in this section general and focus on generic deformations $V(x)$
satisfying the requirement \eqref{pert} for generic $n$ with
$q^n=1$. Later, we will make all of this explicit for the cases $n=2$
with a cubic deformation, and $n>2$ with a quadratic mass deformation.

>From the saddle point equation \eqref{spet} it is
clear that $x$ is a multi-valued function on $\tilde\Sigma$ since it
jumps by a phase factor $q=e^{2\pi i/n}$ at points where the cuts are
identified. However, this
means that $u=x^n$ {\it is\/} single-valued on $\tilde\Sigma$.
So the saddle-point
solution of the matrix model has led us to
a Riemann surface which admits two meromorphic functions $t$ and
$u$. Both $t$ and $u$ have poles at $u=\infty$
and zeros at $u=0$. Furthermore, since $\tilde\Sigma$ is covered once
by the $x$-plane with cuts, it is an $n$-fold cover of the complex
$u$-plane with an $n$-fold branch cut joining all the sheets and
running between $u=0$ and $u=\infty$. There are additional cuts
$\tilde\CC_a$, the image of each
$\CC_a$, which join adjacent pairs of sheets in the cover.

There is a theorem that if $t$ and $u$ are two
meromorphic functions on a Riemann surface then there exists a
polynomial function in two variables such that\footnote{For example,
  in Farkas and Kra \cite{FK} Proposition IV.11.6.}
\EQ{
\tilde F(t,u)=0\ .
}
In the classical limit, where the cuts $\tilde\CC_a$ collapse
into poles we must
recover the singular classical curve $\Sigma$ \eqref{defs}, $F(t,u)=0$. It
is useful to write
\EQ{
\tilde F(t,u)=F(t,u)+\delta(t,u)\ ,
\label{defc}
}
where $\delta(t,u)$ is the quantum correction. This correction
is constrained by
the behaviour of $t$ at $u=0$ and $u=\infty$. Note that due to the
branch points at both these points a good
coordinate is $x$ rather than $u$ itself. Since
$\omega(x)={\cal O}(1/x)$ for large $x$, we have
\EQ{
t\,\underset{x\to\infty}=\,f(x)+{\cal O}(1/x)\ .
}
Correspondingly for small $x$ we have
\EQ{
t\,\underset{x\to0}=\,{\cal O}(x)\ .
}
These two conditions constrain $\delta(t,u)$ to have the form
\EQ{
\delta(t,u)=\sum_{b\geq 0,a\geq1\atop na+\ell b<(n-1)\ell}\gamma_{ab}u^at^b
}
The parameters $\{\gamma_{ab}\}$ are moduli for the solution of the
saddle-point equation. One can count the number of these moduli, they
number precisely the number of singularities $g$ of $\X$ in \eqref{defg};
this is also the (generic) genus of the curve $\tilde\Sigma$.
In other words, there is a modulus for each singularity of the
classical curve $\Sigma$. The picture is now clear. A general solution
to the saddle-point equation involves a deformation of the classical
curve where each singularity at which pairs of the $n$ copies of the
complex $u$-plane touch are resolved into branch cuts. The extent of
each branch cut is then controlled by a single modulus. This is
precisely the quantum geometry advocated in \cite{ber2}.

Notice that each singularity is only associated to a {\it single\/}
cut in the deformed geometry. This means that the matrix model
geometry can only accommodate a single density for each pair of
fractional branes $\RR_\mu^+$ and $\RR^-_\mu$. Otherwise one would expect,
in general, two cuts in the vicinity of each singularity. This
conclusion matches precisely the physics of the RG cascade.
The basic point is the following. Suppose that in the
vacuum of the field theory $N^+_\mu>N^-_\mu$, so there is a
net number of $\RR^+_\mu$ fractional branes. In this case, the low
energy theory after the cascade should involve a $U(N^+_\mu-N^-_\mu)$
gauge group which confines. The matrix model has a single density for
the net number of $\RR^+_\mu$ branes and,
in particular, there is only one glueball field.
Conversely, if $N^+_\mu<N^-_\mu$ then after the RG cascade a
$U(N^-_\mu-N^+_\mu)$
gauge group is left which confines. But once again there is only a
single glueball field.

So the message from the matrix model is in our vacuum we should only
take the net number of fractional branes
\EQ{
N_\mu\equiv N^+_\mu-N^-_\mu
}
to infinity in the matrix model. If $N_\mu>0$  ($<0$) then these
represent $\RR^+_\mu$ ($\RR^-_\mu$) branes. The remaining
$\text{min}(N^+_\mu,N^-_\mu)$ fractional branes come as
$\RR^+_\mu\oplus\RR^-_\mu\simeq \RR^{(n)}[u_\mu,0,0]$
and can move off into the bulk of $\X$. As advocated in
\cite{ber1,ber2}, these branes should be treated as
probes in the matrix model calculation. This makes sense since these
degrees-of-freedom have a moduli space and so we cannot integrate out the
fluctuations around them: they remain as probes which feel the quantum
geometry deformed by the $|N^+_\mu|$ fractional branes. The
important conclusion from \cite{ber1,ber2} is
is that the dynamics of the matrix model can be
considered in isolation: there is no ``back-reaction'' from the
probes. The effect of the fractional branes is to
change the geometry felt by the branes in the bulk.

Given that there will be only one type of fractional brane in the
matrix model from each pair depending upon whether $N^+_\mu\gtrless
N^-_\mu$, we still have to interpret the single cut joining the two
sheets in terms of a density of eigenvalues.
Recall that in the classical limit
the singularity in question is located at $u=u_\mu$.
After the deformation, the singularity
opens up into a cut whose image in the cut $x$-plane involves
two cuts $\CC_\mu$ and $q^{p_\mu}\CC_\mu$. From this we can
reconstruct the density for the fractional. Classically, the
eigenvalues are $q^{j-1}e_\mu$, $j=1,\ldots,p_\mu$ for $\RR^+_\mu$ and
$j=p_\mu+1,\ldots,n$ for $\RR^-_\mu$.
In the matrix model these must spread out onto $p_\mu$ and $n-p_\mu$ cuts,
respectively. Let us suppose $N^+_\mu>N^-_\mu$, then in order to explain
the analytic structure of $t$ it must be that
the $p_\mu$ cuts are rotational translates of the form
$q^{j-1}\CC_\mu$, $j=1,\ldots,p_\mu$. In addition, the densities along
these cuts as measured by the discontinuities in $\omega(x)$
are all equal. This is natural
since these cuts are associated to $\RR^+_\mu$ whose eigenvalues are
$q^{j-1}e_\mu$, $j=1,\ldots,p_\mu$.
This means that
$t$ defined in \eqref{deft} is left with two cuts at $\CC_\mu$ and
$q^{p_\mu}\CC_\mu$ which are then glued
together by the saddle-point equation \eqref{spet}. The whole argument can
repeated in terms of $\RR^-_\mu$ fractional branes if $N^+_\mu<N^-_\mu$.

It follows that if $\AA_\mu$ is a contour surrounding the
cut $\tilde\CC_\mu$ on the curve $\tilde\Sigma$ then
\EQ{
S_\mu\equiv g_s\hat N_\mu=g_s(\hat N^+_{\mu}-\hat
N^-_\mu)=-\frac{1 }{2n\pi}
\oint_{\AA_\mu}\frac{t\,du}u\ .
\label{defts}}
Note that $S_\mu$ is positive (negative) for $N^+_\mu>N^-_\mu$
($N^+_\mu<N^-_\mu$). The quantities $\{\tilde
S_\mu\}$ and $\{\gamma_{ab}\}$ form two bases for the moduli
space of the curve. The quantity $S_\mu$ is interpreted as the
glueball field of the $U(|N_\mu|)$ gauge group factor that confines in
the IR.

The final problem remaining is to fix the moduli of the curve in terms
of the underlying coupling of the theory. This is done by extremizing
the glueball superpotential. The form of this superpotential can be deduced
from earlier work \cite{mm3,mm2,glueball}. Recall that the glueball
superpotential can
only depend on one glueball field for each fractional brane pair. This
motivates the expression
\EQ{
W=\sum_{\mu=1}^gN_\mu\frac{\partial{\cal F}_0}{\partial\tilde
  S_\mu}-2\pi
i\tau\sum_{\mu=1}^g
\hat p_\mu S_\mu\ ,
\label{gbs}
}
where we have defined
\EQ{
\hat p_\mu=\begin{cases}p_\mu & N^+_\mu>N^-_\mu\\
p_\mu-n & N^+_\mu<N^-_\mu\ .\end{cases}
}
Here, ${\cal F}_0$ is the first term, ``genus zero'', in the expansion of
the
free-energy of the matrix model: $\ln{\cal Z}=\sum_{g=0}^\infty{\cal F}_g
g_s^{2g-2}$. We have already seen that $S_\mu$ can be expressed
as an integral around the contour $\AA_\mu$ which surrounds the cut
near $u=u_\mu$. The first term in \eqref{gbs} can also be expressed in
terms of a contour integral on $\tilde\Sigma$. It is the variation of
the genus zero free-energy (times $g_s^{-1}$) upon transporting $|\hat
p_\mu|$ eigenvalues in from infinity to each of the $|\hat
p_\mu|$ cuts of $\omega(x)$ in the $x$-plane. This can be expressed as the
difference of
contour integral of $t\,du/u$ along a contour that runs out from
infinity along the upper sheet of the cut $\tilde\CC_\mu$ minus one that
starts on the lower sheet of the cut $\tilde\CC_\mu$ out to
infinity. These two open contours clearly join to form a closed
contour which can deformed away from infinity. This contour, which we
denote $\BB_\mu$,  is the conjugate cycle to $\AA_\mu$
on $\tilde\Sigma$ that goes down
the cut $\CC_\mu$ back along the lower sheet, up through the branch
cut joining $0$ to $\infty$ and finally out on the upper sheet to join
the starting point. In fact $\{\AA_\mu,\BB_\mu\}$ form a basis for the
one-cycles of $\tilde\Sigma$. Being careful with the normalization,
\EQ{
\frac{\partial{\cal F}_0}{\partial\tilde
  S_\mu}=
-\frac {i }n \oint_{\BB_\mu}\frac{t\,du}u\ .
\label{dfq}
}
and the glueball superpotential is
\EQ{
W=-\frac {i }n\sum_{\mu=1}^g\Big(\tilde
N_\mu\oint_{\BB_\mu}\frac{t\,du}u
-\hat p_\mu\htau\oint_{\AA_\mu}\frac{t\,du}u\Big)\ .
\label{gbs1}
}

Now we consider extremizing $W$ with respect to $S_\mu$.
>From the definition of $S_\mu$
in \eqref{defts}, the one-form
\EQ{
\omega_\mu=-\frac{1 }{2n\pi}\frac{\partial}{\partial\tilde
  S_\mu}\Big(\frac{t\,du}u\Big)\ ,
}
(taken at constant $u$) has the property
\EQ{
\oint_{\AA_\mu}\omega_\nu=\delta_{\mu\nu}\ .
}
Hence, $\{\omega_\mu\}$ is a normalized basis of differentials of the
third kind on $\tilde\Sigma$. The critical equations can then be
written in terms of the period matrix $\tau_{\mu\nu}$ of $\tilde\Sigma$:
\EQ{
\sum_{\mu=1}^gN_\mu
\oint_{\BB_\mu}\omega_\nu=\sum_{\mu=1}^gN_\mu\tau_{\mu\nu}=
\hat p_\nu\htau\ .
\label{cpe}
}
Since $\tilde\Sigma$ has $g$ moduli these $g$ equations are enough to
determine $\tilde\Sigma$ exactly.

The equations \eqref{cpe} are precisely the condition that
$\tilde\Sigma$ is an $\sum_{\mu=1}^g\hat p_\mu N_\mu$-fold
cover
of a torus of complex structure $\htau$. In order to prove this we
need to find a map from $\tilde\Sigma$ to the torus $E_{\htau}$
that covers the latter $\sum_{\mu=1}^g\hat p_\mu N_\mu$ times. For a point
$p\in\tilde\Sigma$, the map is
\EQ{
z(p)=2\pi i\int_{p_0}^p\sum_{\mu=1}^gN_\mu\omega_\mu\text{
  mod }2\pi i,2\pi i\htau\ ,
}
where $p_0$ is an arbitrary base point. In particular, the
glueball superpotential at the critical point is equal to
\SP{
W^*&=-\frac {i }n
\sum_{\mu=1}^g\Big(\oint_{\AA_\mu}dz\,
\oint_{\BB_\mu}\frac{t\,du}u-\oint_{\BB_\mu}dz\,\oint_{\AA_\mu}
\frac{t\,du}u\Big)\\
&=-2\pi \,{\rm Res}_{\infty}\frac{f(x)z\,dx}x\ .
\label{mmre}
}
In the above, we applied
a Riemann bilinear relation and used the fact
that $dz=\sum_\mu N_\mu\omega_\mu$ and that $t$ can be replaced with
$f(x)$ in the vicinity of its pole at $u=\infty$ ($x=\infty$).

One remaining issue is to
include the bulk branes as probes in the
matrix model calculation and to work out the effect of the fractional
branes on these probes. We will simply quote the results of \cite{ber2}
here.
The
classical moduli space of a bulk brane $\X$ is modified by simply
replacing the function
$f(x)$ by $t(x)$. The second term in \eqref{deft} represents the effect of
the fractional branes on the geometry.
In other words we replace in \eqref{deformed}
$F(t,u)$ by the deformation $\tilde F(t,u)=F(t,u)+\delta(t,u)$.
This defines a quantum deformed moduli space space $\tilde\X$:
\EQ{
uvw=(-1)^{n+1}\tilde F(t,u)\ .
}

\subsection{The case $q=-1$ with cubic deformation}

As we have noted earlier the theory with $q=-1$ and cubic
superpotential \eqref{restricted} exhibits a duality cascade in a
certain regime of
parameter space. This example will allow us
to explicitly see in a simple setting, many of the general features
argued above. There is one pair of fractional branes so that the deformed
curve $\tilde\Sigma$ is a torus. Its form will be determined by a
deformation of the classical curve (following from \eqref{CYdef})
\EQ{F(t,u)=t^2-\frac{\epsilon^2}4u(u-a^2)^2=0\ .
}

In this particular case, we can also derive the quantum deformed curve
directly from the matrix model
loop equations and since this is interesting we
pause to do this.  The loop equations can be derived by
noting that the matrix integral is invariant under generic
reparameterizations. In particular consider a variable shift $\delta
x_i \sim 1/(x-x_i)$ which should leave invariant
the integral Eq.~\eqref{eigenintegral}. Following the general algebraic
manipulations shown in Appendix A we are led to the following
condition (for any finite $\hat N$):
\EQ{\langle t(x)^2\rangle=\Big\langle\tfrac14x^2\;V^\prime(x)^2 +
g_s \sum_i x^2\;(V^\prime(x_i)
-V^\prime(x))\Big[{1\over x-x_i} -{1\over
    x+x_i}\Big]\Big\rangle\ .\label{quad}}
In the large-$\hat N$ limit, when all averages are dominated by the
saddle-point configuration of the matrix integral, we can treat this
equation
as an ordinary algebraic equation for $t$.
The first point to note here is that, when $V(x)$ is an odd
polynomial, $t(x)$ is completely
determined by a polynomial function on the right hand side of
Eq.~\eqref{quad}.
(If $V(x)$ includes
even powers, this is no longer true; the right hand side is
non-analytic and one also needs to consider higher loop
equations.) But we are interested precisely in the cubic polynomial
deformations Eq.~\eqref{restricted}, and in the large-$\hat N$ limit
\EQ{
t^2= {1\over 4}\epsilon^2\;x^2(x^2-a^2)^2 + 2\;x^2\;\epsilon
  \int_\CC\rho(x)\,dx\
.\label{CYmm}
}
where on the right hand side we encounter a constant (times $x^2$) which is
the
first moment of the eigenvalue distribution. Now we can clearly
see how this is related to the deformed (classical) geometry
Eq.~\eqref{CYdef}
which was the classical moduli space of the field theory. Identifying
the variable $u$ in Eq.~\eqref{CYdef} with $x^2$ in Eq.~\eqref{CYmm},
the above
equation reduces to
\EQ{\tilde F(t, u)=t^2 -{\epsilon^2\over
    4}u(u-a^2)^2+\gamma u=0\label{loopeq}\ .}
where $\gamma$ is the constant determined by the first moment of the
eigenvalue distribution. This is precisely the quantum deformed curve
\eqref{defc}. The effect of the deformation is to split the double
zero at $u=a^2$.
Thus the function $t$ lives on the two-sheeted complex
$u$-plane with two branch cuts. The associated Riemann surface is a
torus. It now remains to show how the deformation $\gamma$
will depend on gauge theory parameters. This can be
obtained directly by expressing the hyperelliptic curve
\eqref{loopeq} above in Weierstrass form as we show below.

It is also instructive to see how all this follows from the saddle
point equation \eqref{spet} for $t(x)$. In this case $t(x)$ has two
branch cuts, centred at $x=\pm a$ and is defined as
\EQ{t(x)= -{1\over 2}\epsilon \;x
  (x^2-a^2)-S[x\omega(x)+x\omega(-x)].\label{deft1}}
This in turn implies that $t(x)=-t(-x)$ and hence the two branch cuts
are images of each other. The saddle point equation
provides a gluing condition on these two cuts as illustrated in
Fig.~1.
\begin{figure}
\begin{center}\mbox{\epsfbox{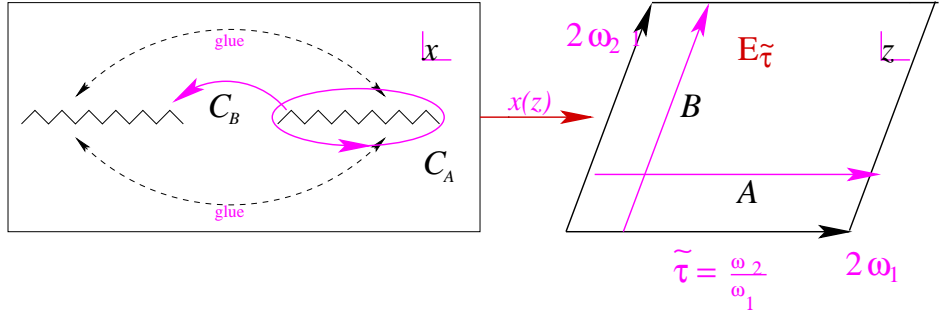}}\end{center}
\caption{\small The branch cuts of $t(x)$ and the gluing conditions on
  them lead to a parameterization in terms of a torus $E_{\tilde \tau}$
  with complex structure parameter $\tilde\tau$.}
\end{figure}
The gluing conditions on the two cuts imply that we should think of the
Riemann surface as the complex plane with a handle which is a
torus. To see this torus explicitly and connect it to
gauge theory parameters we first look for an elliptic or Weierstrass
parameterization of the problem. (This is exactly the problem that was
solved in \cite{mm3}.) In this
parameterization the torus of complex structure
$\tilde\tau$, which we denote $E_{\tilde\tau}$, is realized as the
fundamental parallelogram in the complex $z$-plane
with sides $2\omega_1$ and
$2\omega_2$ with opposite identification. Its
complex structure is $\tilde \tau=\omega_2/\omega_1$. There is a
natural map $x(z)$ from the torus $E_{\tilde{\tau}}$ to the two-cut
complex $x$-plane, which must satisfy the conditions:
\AL{&A-{\rm cycle}\;{\rm shift}:\quad z\rightarrow
  z+2\omega_1:\quad x(z)\rightarrow x(z);\quad
  t(x(z))\rightarrow t(x(z))\ ,\\
&B-{\rm cycle}\;{\rm shift}:\quad z\rightarrow
  z+2\omega_2:\quad x(z)\rightarrow - x(z);\quad
  t(x(z))\rightarrow t(x(z))\ .}
Hence $x(z)$ is quasi-elliptic, while $t(x(z))$ is an elliptic function
on the torus $E_{\tilde{\tau}}$. Both these functions can be
determined precisely in terms of the torus variables, from the
(quasi)periodicity conditions above and the large-$x$ asymptotics
determined by Eq.~\eqref{deft1}.
We find,
\EQ{x(z)= A
\sqrt{\;\wp(z) \;- \;e_1(\tilde\tau)}\;\;;\quad t=
-{\epsilon\over 4}\;A^3\;\wp^\prime(z)\ ;\quad A=-ia\sqrt{2\over
  3}{\;[e_1(\tilde\tau)]^{-{1\over 2}}} \label{xelliptic}\ .}
Here we have used standard
notation, $e_i(\tilde\tau)=\wp (\omega_i)$ is the
Weierstrass function evaluated at one of the three half-periods
$\omega_1,\omega_2$ and
$\omega_3=\omega_1+\omega_2$.
The constants of proportionality
are uniquely determined by the large-$x$ asymptotics
required by Eq.~\eqref{deft1}, which maps to the behaviour in the vicinity
of the pole at $z=0$.
Now, since
\EQ{u=x^2(z)=A^2\big(\wp(z)-e_1\big),}
as anticipated in our general arguments in the previous
section, $u$ and $t$ will satisfy an algebraic relation which in this
case follows
from the basic differential equation satisfied by the Weierstrass
function,
\EQ{
{\wp^\prime}^2(z)=4(\wp(z)-e_1)(\wp(z)-e_2)(\wp(z)-e_3)\ .\label{wsf}}
This leads precisely to the algebraic equation \eqref{deft}
with
\EQ{
\gamma=\frac{\epsilon^2a^4}{36}
\left[\frac{e_3(\ttau)-e_2(\ttau)}{
    e_1(\ttau)}\right]^2\ .
\label{gamma}}
Indeed, we could have deduced this directly from Eq.~\eqref{loopeq} by
going to an elliptic parameterization, but it is useful to see how the
loop equation approach matches up with the saddle point approach.

The relation between the modular parameter $\ttau$ of the torus to
gauge theory parameters emerges upon extremizing the glueball
superpotential of the theory. The glueball superpotential depends on
the following period integrals \eqref{defts} which can be explicitly
evaluated in terms of the modular parameter $\ttau$
\EQ{S = -{1\over 4\pi }\oint_{\AA} {t du\over u}
=\frac1{2\pi i}{dh(\ttau)\over
  d\ttau}}
and
\EQ{{\partial {\cal F}_0\over \partial S}=-{i \over 2}
  \oint_{\cal B} {t du\over u}=\ttau {dh(\ttau)\over d\ttau}-h(\ttau)}
where
\EQ{
h(\ttau)= \epsilon a^3\sqrt{\frac2{27e_1(\ttau)}}\ .
}
As we saw from our general arguments, extremizing the glueball
superpotential Eq.~\eqref{gbs}, written explicitly in terms of the
expressions above, fixes the complex structure parameter of the torus
in terms of the gauge theory coupling in a simple way
\EQ{
\tilde\tau=\frac{\htau}{|N^+-N^-|}\ .\label{ttau}}
This is what we obtain from \eqref{cpe} by noting that the single
element of the period matrix of a torus is its complex structure
parameter.

Finally we have the critical value of the glueball
superpotential, which can be determined explicitly using the above
expressions or directly from Eq.~\eqref{mmre}
\EQ{W^*=|N^+-N^-| \epsilon
a^3\sqrt{2\over 27e_1(\ttau)}\ .
}
where $\ttau=\htau/|N^+-N^-|$. The superpotential has a nontrivial
expansion in (fractional) instantons when ${\rm Im}(\ttau)\gg1$ and
vanishes when $N^+=N^-$. This is in line with our expectation that in
the absence of fractional branes the low-energy theory is the 
Klebanov-Witten
conifold SCFT. Note also that the deformation parameter $\gamma$
also vanishes in this case (this follows from the fact that
both $e_2(\ttau)$ and $e_3(\ttau)$ approach the same constant
value in the limit ${\rm Im}(\ttau)\rightarrow \infty$). An
important feature of the IR gauge theory is reflected in this critical
superpotential, namely that it is not pure $\N=1$
SYM. It has massive modes which are not decoupled from
the $\N=1$ gauge multiplet. The theory does reduce to the pure $\N=1$
gauge theory in a decoupling limit $\hat \tau\rightarrow i\infty$ keeping
fixed
the dynamical scale $a\epsilon^{1/3}\;\exp(2\pi i\ttau/3)$. This is also
apparent in the structure of the deformed geometry.

The full deformed Calabi-Yau geometry (Fig. 2)
\begin{figure}
\begin{center}\mbox{\epsfbox{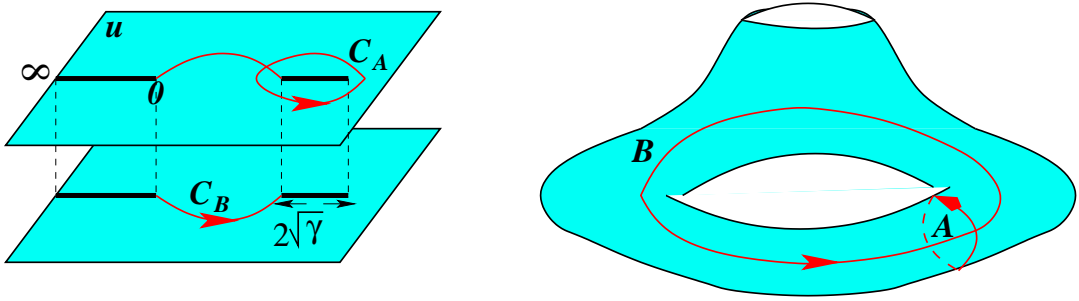}}\end{center}
\caption{\small The one-cycles on the $u$-plane which lift to the
  compact 3-cycles of the Calabi-Yau geometry.}
\end{figure} is obtained from the matrix
model Riemann surface adding $uvw$ to the right hand side of
Eq.~\eqref{loopeq}.
In this deformed Calabi-Yau geometry:
\EQ{t^2 = -uvw +{\epsilon^2\over 4}u(u-a^2)^2+ u \gamma}
the isolated conifold singularity at $u=a^2$ has been replaced by a
3-sphere of a finite size controlled by $\gamma$ which in turn is
determined in terms of gauge theory parameters by Eqs. \eqref{gamma}
and \eqref{ttau}. This
is the deformation seen in the geometric dual at the endpoint of the
cascade.

It is also interesting to compare and contrast the loop
equation (the matrix model Riemann surface) for the $\hat A_1$ quiver
theory
and that for our theory with $q=-1$. Recalling the result of
\cite{dvdecon}, the infrared physics at the endpoint of the cascade in
the $\hat A_1$ quiver theory is controlled
by a matrix model curve:
\EQ{\qquad y^2= x^2+\mu^2,}
which is the complex $x$-plane with a single branch cut. There are two
one-cycles, one of which
is compact (the contour enclosing the cut) while its dual cycle is
noncompact (running from a branch point to infinity). This
lifts to the deformed conifold geometry $y^2=uv+x^2+\mu$ with one
three-cycle of finite size. This is, of course, what one
expects for the pure $\N=1$ SUSY Yang-Mills theory in the IR.

In contrast, as we have seen in Eq.~\eqref{loopeq} the theory under
investigation leads to the two-cut complex $u$-plane with {\em two\/} 
compact
cycles. One of these corresponds to the contour enclosing the cut
between $u=a^2+2i\sqrt\gamma/\epsilon$ and
$u=a^2-2i\sqrt\gamma/\epsilon$. The dual cycle is given by the
contour
running from $u=0$ to $u=a^2+2i\sqrt\gamma/\epsilon$, moving down to the
second sheet and running back to $u=0$ (which is the tip
of the second cut extending
from $u=0$ to $u=\infty$).
In a certain limit $(a\rightarrow\infty)$ where one of
these cycles becomes noncompact we recover the physics of the
pure $\N=1$ SYM in the  infrared. This is also the decoupling limit
for the adjoint fields in Eq.~\eqref{supeff}.

\subsection{The case $n>2$ with quadratic deformation}

This case with quadratic superpotential \eqref{mass} allows us to
embed the duality cascade within a relevant deformation of the
UV-finite four-dimensional SCFT. We will now determine the
quantum deformed geometry associated to this theory. Recall that
classically, the Higgs branch moduli space of this theory is described
by the algebraic curve Eq.~\eqref{deformed}, $uvw=(-1)^{n+1}F(t,u)$. It has
$[n/2]$ isolated singularities of the conifold type where $v=w=0$
and where the (classical) Riemann surface described by $F(t,u)=0$
becomes singular. Here the classical curve
\EQ{F(t,u)=\prod_{i=1}^n\Big(t+\epsilon\frac{q^{2i}x^2}{1-q^2}-\epsilon
a\frac{q^ix}{1-q}\big)=0\ .}
Note that this curve is only a function of $u=x^n$ since it is
invariant under $x\rightarrow xq$. The singularities occur at the
points $u=u^{(p)}$, $p< n$ given by Eq.~\eqref{singularities} where the
classical curve above has a double zero. Fractional branes can be
located at each of these conifold
singularities.
In all there are $[n/2]$ distinct types of fractional brane pairs. The
effect of sticking fractional branes at a conifold singularity is to
split the double zeroes, resulting in a branch cut that
joins a pair of $n$ sheets of the Riemann surface associated to $t$ as a
function of $u$. This is exactly what we encountered in the previous
example with $n=2$. This deformation leads us to the quantum curve
\eqref{defc} which takes the form
\EQ{
\tilde
F(t,u)=\prod_{i=1}^n\Big(t+\epsilon\frac{q^{2i}x^2}{1-q^2}-\epsilon
a\frac{q^ix}{1-q}\big)+u\sum_{a=0}^{[(n-3)/2]}\gamma_at^a=0\ .
}
The $\gamma_a$ are the quantum deformation parameters which we would
like to determine using the matrix model. They are implicitly functions
of the glueball condensate of the field theory.
The curve is actually hyper-elliptic since it has the general form
\EQ{
\tilde
F(t,u)=t^n+\Big(\frac\epsilon{1-q^2}\Big)^nu^2+u\sum_{a=0}^{[n/2]}
s_at^a=0\ ,\label{qcurve}}
where the $s_a$ depend on the moduli. In fact, $s_a$ is equal to
$\gamma_a$, for $a=0,\ldots,[(n-3)/2]$, up to a constant shift, while
$s_a$, for $a=[(n-1)/2],\ldots,[n/2]$ only depend on the parameters in
the potential.

Generically when all types of fractional branes are
present the curve has genus $[n/2]$, as each of the singularities
gets opened up into a branch cut. However,
let us consider the simplest vacuum, which was described in Section
3.2 where fractional branes are present at only one of the conifold
singularities.
This means that only one cut opens up leaving a genus one Riemann
surface, or torus. In this case, we can solve for $t$ and $u$
(and $x$) in terms of the Weierstrass parameterization.

The parameterization we seek can be deduced from the saddle-point
condition \eqref{spet}, which says that the two cuts of $t(x)$ which are
images under a rotation by $q$ are now glued together. This adds a
handle to the complex $x$-plane giving us a torus (Fig.~2) with a
complex structure modulus $\ttau$. As explained in Section 4.3,
we think of this torus
$E_\ttau$ as the complex $z$-plane modded out by lattice
translations. Going around the A-cycle of
this torus transforms $x$ by a phase $x(z)\rightarrow q x(z)$ while the
B-cycle
shift leaves $x(z)$ invariant. Clearly this means that the
$u(z)=x^n(z)$ is an elliptic function of $z$ (this is true
only because $q$ is an $n$-th root of unity). These operations also
leave $t(x(z))$ invariant. Along with the large-$x$ asymptotics (which
maps to small-$z$), this
determines the two functions uniquely:
\EQ{u(z)=
A^n\cdot{\theta_1^n({\pi z/2\omega_1}-p{\pi/n})\over
\theta_1^n({\pi z/2\omega_1})}\ ;\qquad
t(x(z))=
B\cdot[\wp(z)-\wp(2\omega_1 p/n)]\ .\label{elliptic}}
where the constants $A$ and $B$ are
\EQ{A=a\cdot{(1+q)\over
    2}\cdot{\theta_1^\prime(0)\over\theta_1^\prime({p\pi/n})}\ ;\qquad
B= -\epsilon a^2\cdot{\omega_1^2\over \pi^2}\cdot{1+q\over 1-q}\cdot
{\theta_1^2(p\pi/n)\over\theta_1^{\prime 2}(p\pi/n)}}
The modular parameter associated to all the (quasi)elliptic functions
appearing in the above expressions is $\ttau$, the complex structure
parameter of the torus $E_\ttau$.

$u(z)$ is an elliptic function with an $n$-th order pole at
$z=0 $ and an $n$-th order zero at $z=2p\omega_1/n$. By determining
the coefficients of the singular (and constant) pieces in the Laurent
expansion of $u$ around $z=0$, we can rewrite $u(z)$ in terms of the
Weierstrass function and its derivatives using well-known results for
elliptic functions (Eq.~\eqref{repweier}). This straightforward (but 
lengthy)
procedure allows us to obtain the
algebraic relation between $u$ and $t$. For example for $n=3$  and
$p=1$ we find
\SP{\tilde F(t,u)&=
t^3+\Big[{\epsilon\over{1-q^2}}\Big]^3\;u^2 -
3a\Big[{\epsilon\over(1-q^2)}\Big]^2\; tu\\ &
\qquad\qquad + {a^3\over 8}
\Big[{\epsilon\over{1-q}}\Big]^3 {\theta_1^3(\pi/3)\over
  \theta_1^{\prime 3}(\pi/3)}\;\wp^\prime(2\omega_1/3)\;u=0\ .}
This quantum curve has all the properties that we anticipated in and
below Eq.~\eqref{qcurve}. (For the $n=3$ case, where $[n/2]=1$, there is
actually only one conifold singularity.) In particular there is only
one quantum
deformation parameter which alters the coefficient of the term linear
in $u$. As we
see below the classical limit---absence of fractional
branes---corresponds to taking ${\rm Im}(\ttau)$ to $\infty $. In this limit
the quantum deformation
that we have computed above indeed approaches the
result expected from the classical curve Eq.~\eqref{deformed}. The
classical Riemann surface associated to $t(u)$ consists of three
sheets joined via the branch cut running from $u=0$ to $u=\infty$. The
effect of the quantum deformation is to open up the singularity at
$u=a^3$ into a square root branch cut that joins two of the three
sheets (Fig.~3).
\begin{figure}
\begin{center}\mbox{\epsfbox{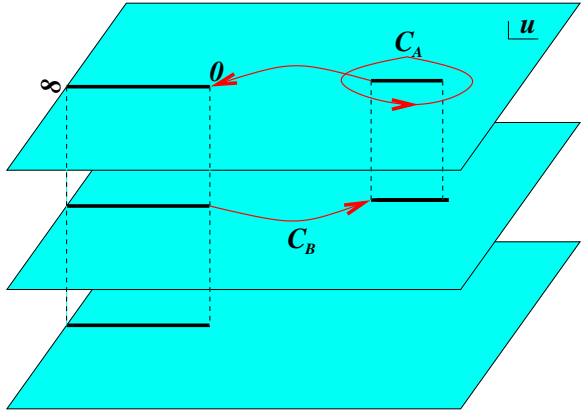}}\end{center}
\caption{\small The Riemann surface for $t(u)$ in the case $n=3$. The
  two one-cycles on the $u$-plane lift to
  compact 3-cycles of the Calabi-Yau space.}
\end{figure}
The algebraic equation for the deformed Calabi-Yau space is, as before
\EQ{uvw=(-1)^{n+1}\tilde F(t,u).}


Once again we can relate the complex structure $\tilde\tau$ to the
gauge theory coupling by extremizing the glueball superpotential as in
\eqref{cpe}. For a gauge theory vacuum where we have $N^+$ and $N^-$
fractional branes of the two different types, we find the
relation\footnote{This can be explicitly seen by evaluating $S$
  and $\partial {\cal F}_0/\partial S$ which turn out to be
  $dh(\ttau)/d\ttau$ and $\ttau
  dh(\ttau)/d\ttau-h(\ttau)$ where $h(\ttau)=2 \epsilon a^2
  [(1+q)/(1-q)] \theta_1(p \pi/n)/\theta_1^\prime(p
  \pi/n)$. Extremizing the superpotential \eqref{gbs} then leads to
  the desired relation between $\ttau$ and the gauge theory parameter.}
\EQ{
\tilde\tau=
\frac{\hat p\htau}{N^+ -N^-}=\begin{cases}
p\htau/(N^+-N^-) & N^+>N^-\\
(n-p)\htau/(N^--N^+) & N^+<N^-\ .\end{cases}
}
Hence, the limit where the deformation vanishes $i.e.$ $N^+=N^-$ is
  when ${\rm Im}(\ttau)\rightarrow \infty$.
Finally from \eqref{mmre} we have the critical value of the glueball
superpotential:
\EQ{
W^*=-2|N^+-N^-| \epsilon a^2{1+q\over 1-q}.
\;{\theta_1(p\pi/n)\over\theta_1^{\prime }(p\pi/n)}
}
where the modular parameter of the Jacobi theta functions, $\ttau$ is
  determined as above. This value of the effective superpotential
coincides with that of the mass deformation of the $\beta$-deformed
  theory obtained in \cite{mm3}. This indicates that at least at the
  level of the holomorphic sector, the low-energy theory at the end of
  the RG cascade is the mass deformation of the $U(|N^+-N^-|)$
  $\beta$-deformed theory (with $\beta=2\pi/n$). Pure $\N=1$ SUSY
  gauge theory only emerges in a decoupling limit for the extra
  massive adjoint fields.

\subsection{Relation to periods of the Calabi-Yau geometry}

It is interesting to explicitly see the relation of the matrix
model moduli to periods of the Calabi-Yau geometry. We know that a
vacuum of the $\N=1$ theory with
$U(N^+)\times U(N^-)$ gauge symmetry
is obtained by taking $N^+$ fractional D3-branes of one type and $N^-$
of the second type and placing them at a conifold singularity
in the (classical) geometry specified by Eq.~\eqref{deformed}.
The general picture is that gaugino
condensation in the infrared theory leads to a geometric dual where
the $S^2$ is
replaced by a deformed $S^3$ without any branes. For the B-model
topological string in this deformed geometry, the genus zero
prepotential term is determined by period integrals of the holomorphic
$(3,0)$ form. These period integrals of the deformed Calabi-Yau
geometry should precisely be the matrix model integrals of the
meromorphic one-form over the $\AA$ and
$\BB$ cycles of the Riemann surface.  We show that this is indeed
the case.

First, in the simplest case with $q=-1$, consider the holomorphic
$(3,0)$ form associated to the geometry \eqref{CYdef},
\SP{\Omega&=dt^\prime\wedge du\wedge dv \wedge
  dw\;\delta(\;t^{\prime 2}+uvw-{\epsilon^2\over 4}u(u-a^2)^2+\gamma
  u)\\
&={du\wedge dv \wedge dw\over2\sqrt{u\left[-vw+
{\epsilon^2\over 4}(u-a^2)^2+\gamma\right]} }.}
The geometry has compact 3-cycles over which the period integrals can
be evaluated as follows. First one picks an appropriate reality
condition on the
coordinates ({\it e.g.} $v=w^*$ assuming that $a$ and $\gamma$ are
real, positive). The 3-cycles can be identified by treating $v$ and
$w$ as coordinates on an $S^2$ which is
fibred over an appropriately chosen interval on the $u$-plane.
In this
way the period integrals over the compact 3-cycles in the Calabi-Yau,
reduce to contour integrals over the branch cuts (Fig.~3) of a meromorphic
1-form in the $u$-plane,
\EQ{\oint_{A,B}\Omega\;= \;2\pi\oint_{{\AA},{\BB}}\sqrt{{\epsilon^2\over
      4}(u-a^2)^2+\gamma}\;{du\over \sqrt{u}} = 2\pi \oint_{{\AA},
    {\BB}}
t\; {du\over u}, }
which are precisely the moduli of the matrix model solution discussed
above.

More generally, we can present an argument along the lines of
\cite{Aganagic:2003qj}. For the non-compact Calabi-Yau given as the
hypersurface in ${\bf C}^4$,
\EQ{uvw =(-1)^{n+1}\tilde F(t,u)}
where $\tilde F(t,u)$ is defined as in Eq.\eqref{defc}, we think of
the Calabi-Yau space $\bf X$ as a fibration over the $(t,u)$-plane,
where the fibre is the curve $uvw=(-1)^{n+1}\tilde F$. The
holomorphic $(3,0)$ form can be chosen to be
\EQ{\Omega = {dt\wedge du\wedge dv\over uv}.}
This has periods over the three-cycles in $\bf X$ which reduce (by Cauchy's
theorem) to
\EQ{\int_D {dt \wedge du\over u}}
where $D$ is a real two-dimensional domain in the complex $(t,u)$-plane
such that $\partial D\subset \tilde \Sigma$. Here $\tilde\Sigma$ is the
Riemann surface $\tilde F(t,u)=0$ on which the fibre degenerates. By
Stokes' theorem these integrals reduce to
\EQ{\oint_{\AA_\mu, \BB_\mu}{t du\over u}}
integrals of a meromorphic one-form over the one-cycles of the Riemann
surface $\tilde \Sigma$. These are precisely the moduli $(\tilde
S_\mu, \partial {\cal F}_0/\partial\tilde S_\mu)$ of the matrix
model (Eqs.\eqref{defts} and \eqref{dfq}.)
\section{A Generalization}

Finally, we turn to a more general class of renormalizable deformations of 
the
superconformal field theory (with $q^n=1$) described by the superpotential
\EQ{
W= \Tr[XYZ - q XZY-m_1YZ-m_2XZ-m_3XY]\ .
\label{pertn}
}
By shifting each field by a multiple of the identity
one has the equivalent description
\EQ{
W= \Tr[XYZ - q XZY-2\sigma X-2\zeta Y-2\eta Z]\ ,
}
where
\EQ{
\sigma=\frac{m_2m_3}{2(q-1)}\,
\qquad\zeta=\frac{m_1m_3}{2(q-1)}\ ,\qquad\eta=\frac{m_1m_2}{2(q-1)}\ .
}
This is precisely the deformation considered in \cite{Douglas2,LB}. It
completely resolves the orbifold fixed lines leaving behind $(n-1)$
isolated conifold singularities.

As previously it is useful to consider a slightly more general form of
the theory by introducing a general superpotential for $X$:
\EQ{
W= \Tr[XYZ - q XZY-2\zeta Y-2\eta Z-V(X)]\ ,
\label{pertnn}
}
We are going to show that this more general deformation
can easily be incorporated into our existing
formalism by a simple trick. The idea is to ``complete the square'',
in $Y$ and $Z$ by defining
\EQ{
Y=\tilde Y-\frac{2\eta}{q-1}X^{-1}\ ,\qquad
Z=\tilde Z-\frac{2\zeta}{q-1}X^{-1}\ .
}
This move changes the superpotential for $X$ to
\EQ{
\tilde V(x)=\sum_{j\neq 0\text{ mod }n}^\ell {a_j\over j} x^{j}
-\frac{4\zeta\eta}{q-1}x^{-1}\ .
}

Our previous formalism now applies with $V$ replaced by $\tilde
V$. Hence the classical moduli space $\X$ of the Higgs branch is now
\EQ{
u\tilde v\tilde w=(-1)^{n+1}F_{\tilde V}(t,u)\ ,
}
where $F_{\tilde V}(t,u)$ is given as in \eqref{deformed} but with
$f(x)$ defined as in \eqref{deff} with $V(x)$ replaced by $\tilde V(x)$ and
\EQ{
\tilde v=\frac1n\Tr \tilde Y^n\ ,\qquad\tilde w=\frac1n\Tr\tilde Z^n\
.
}
Since
\EQ{
v=\tilde v+ \Big(\frac{2\eta}{1-q}\Big)^n\frac1u\ ,\qquad
w=\tilde w+\Big(\frac{2\zeta}{1-q}\Big)^n\frac1u\ ,
}
we can write the moduli space in terms of the original variables as
\EQ{
uvw=(-1)^{n+1}F_{\tilde V}(t,u)+\Big(\frac{2\eta}{1-q}\Big)^nv
+\Big(\frac{2\zeta}{1-q}\Big)^nw-\Big(\frac{2\zeta\eta}{(1-q)^2}\Big)^n
\frac1u\ .
}
It turns out that the right-hand side is a polynomial in $t$ and $u$ since 
the
${\cal O}(1/u)$ terms cancel.

The classical singularity structure of $\X$ is somewhat different from
the case with $\zeta=\eta=0$. The singular line at $u=0$ is now
resolved and only isolated conifold-like singularities
remain. In this case the Higgs branch becomes disconnected from the
Coulomb branch. As before, the
isolated singularities are associated to pairs of
representations $\RR_\mu^\pm$. The matrix model analysis can be
followed as in Section 4 to find the quantum deformed geometry.
The only difference is that the quantum
deformation of the curve is slightly more general with \eqref{defc}
replaced by
\EQ{
\delta(t,u)=\sum_{b\geq 0,a\geq0\atop na+\ell b<(n-1)\ell}\gamma_{ab}u^at^b
}

For example, in the case with $q=-1$ and with $V(x)=2\sigma x$, the
shifted potential is
\EQ{
\tilde V(x)=2\sigma x+\frac{2\zeta\eta}{x}\ .
}
The classical moduli space \eqref{deformed} is
\EQ{
uvw=-t^2+\sigma^2u+\zeta^2v+\eta^2w-2\sigma\zeta\eta\ ,
\label{ndcm}
}
in particular, there is a
single singularity at $v=w=0$ and $u=\zeta\eta/\sigma$.
The quantum curve $\tilde\Sigma$ is
\EQ{
\tilde F(t,u)=t^2-\sigma^2 u-\zeta^2\eta^2\frac1u+
2\sigma\zeta\eta+\gamma=0\
,
\label{ndc}
}
where $\gamma$ is the quantum deformation. This curve is, as expected,
a torus. It can be written in Weierstrass form \eqref{wsf} by taking
\EQ{
u=B\wp(z)\ ,\qquad t=\frac{A\wp'(z)}{B\wp(z)}\ ,
}
where
\EQ{
A=\Big(\frac{g_3}{\zeta^2\eta^2}\Big)^{1/2}\ ,\qquad
B=\Big(\frac{4\zeta^2\eta^2}{\sigma g_3}\Big)^{1/3}\ .
}
Here, $g_2$ and $g_3$ are the Weierstrass
{\it invariants\/} defined in \eqref{winv}.
The quantum deformation in \eqref{ndc} is then
\EQ{
\gamma=g_2\Big(\frac{\sigma\zeta^4\eta^4}{4
  g_3^2}\Big)^{1/3}-2\sigma\zeta\eta\ .
}

The critical point of the glueball superpotential will set the complex
structure of the torus $\tilde\tau$ as in \eqref{ttau}. Replacing $F$ by
$\tilde F$ in \eqref{ndcm} then gives the quantum deformed geometry
$\tilde\X$ as a function of the underlying couplings.

\section{Conclusions}

We have studied relevant deformations of the $\N=1$ SCFT related to $\N=4$
theory by the exactly marginal $\beta$-deformation. We have seen how at
intermediate energy scales, certain vacua of the theory (at weak gauge
coupling) can have a
description in terms of an $\N=1$ theory undergoing a
Klebanov-Strassler duality cascade which terminates in an IR theory with
a mass gap and gaugino condensation. We understood all of this
rather precisely in the language of the Dijkgraaf-Vafa matrix model
which allowed us to compute the quantum deformed moduli space seen by
a probe D-brane at the end of the cascade. Specifically we were able
to compute exactly the deformation parameters as functions of the
microscopic parameters of the gauge theory.
A natural question arising
in the context of these theories, which we hope to address in
future work, is what is the supergravity/string theory dual for these
theories. In the UV the field theory (at large-$N$) is (super)conformal
and is dual
to the near
horizon geometry $AdS_5\times S^5/\Gamma$ ($\Gamma$ being the orbifold
action) of D-branes at the $\C^3/\Z_n\times\Z_n$ orbifold
with discrete torsion.\footnote{When $n$ is large and $q$ is close to
  $1$, we can also think of this theory as a small deformation of the
  $\N=4$ theory and consequently its SUGRA dual corresponds to
  switching on a non-normalizable supergravity mode in the
  $AdS_5\times S^5$ background. The relation between the two
  descriptions has been discussed in \cite{LB1}.}
The discrete torsion should be encoded in the
boundary conditions for massless twisted states of the orbifold. As noted in
\cite{ber2} the relevant deformations we have studied correspond to
turning on certain fields in the twisted sector. This deformation
should cause the $AdS$ geometry to
smoothly match on to a Klebanov-Strassler type geometry that describes
the cascade phenomenon in the IR field theory. It would be extremely
interesting to obtain such a SUGRA flow. Of course, from the point of
view of the field theory results we should also expect to find a flow
interpolating between $AdS_5\times S^5/\Gamma$ and $AdS_5\times
T^{(1,1)}$ which is the conifold theory of \cite{KW}.

{\bf Acknowledgements:} We would like to thank N. Dorey, G. Policastro
and F. Quevedo for discussions.

\startappendix
\Appendix{Matrix model loop equations}
We now outline how the quadratic loop equations follow from the
integral \eqref{eigenintegral}. Performing a variable change $\delta
x_i={\epsilon\over{x-x_i}}$, the invariance of the integral under
general variable changes leads to the condition
\SP{&\langle \sum_i{1\over (x-x_i)^2}+\sum_{i\neq j}
{1\over(x_i-x_j)}\Big[{1\over x-x_i}-{1\over x-x_j}\Big]-
\sum_{i\neq j}
{1\over(x_i-q\;x_j)}\Big[{1\over x-x_i}-{q\over x-x_j}\Big]\rangle\\
&=\langle{\hat N\over S}\sum_i
{V^\prime(x_i)\over{x-x_i}}\rangle.
\label{variation}}
Now, the identities
\EQ{{1\over(x-x_i)(x_i-q\;x_j)}={1\over(x-x_i)(x-q\;x_j)}
+{1\over(x-q\;x_j)(x_i-q\;x_j)}}
and
\EQ{{1\over(x-x_i)(q\;x_i-x_j)}={1\over(x-x_i)(q\;x-x_j)}
+{q\over(q\;x-x_j)(q\;x_i-x_j)}}
can be used recursively along with Eq.\eqref{variation} to obtain the
following condition on the resolvent function
\EQ{\sum_{k=0}^{n-1}\langle
\Big[q^k\omega(xq^k)\rangle -q^{k+1}\omega(x q^{k+1})\Big]^2\rangle
={2\over S}\;{1\over \hat N}\sum_{i=1}^{\hat N}\sum_{k=0}^{n-1}
q^{2k}\langle {V^\prime(x_i)\over x q^k -x_i}\rangle.}
We can rewrite this as
\SP{&\sum_{k=0}^{n-1}\langle
\Big[q^kx\omega(xq^k)-q^{k+1}x\omega(x q^{k+1})+{1\over
  S}f(q^{k+1}x)\Big]^2\rangle
={1\over S^2}\sum_{k=0}^{n-1}f^2(q^{k+1}x)+\\
&+{2\over S} {1\over \hat N}\sum_{i=1}^{\hat
  N}\sum_{k=0}^{n-1}
q^{2k}x^2\langle {V^\prime(x_i)-V^\prime(xq^k)\over x q^k
  -x_i}\rangle.\label{loop1}}
In terms of the function $t(x)$ defined in Eq.\eqref{deft} the
quadratic loop equation simplifies
\EQ{\sum_{k=1}^{n}\langle t^2(q^k x)\rangle=\sum_{k=1}^n f^2(q^k x)
+{S} {2\over \hat N}\sum_{i=1}^{\hat
  N}\sum_{k=0}^{n-1}
q^{2k}x^2\langle {V^\prime(x_i)-V^\prime(xq^k)\over x q^k -x_i}\rangle.}
For $n=2$ and a perturbation that satisfies the condition
\eqref{restricted} the right hand side is a polynomial and the
quadratic loop equation is sufficient to determine $t(x)$ in the
large-$\hat N$ limit when the loop equation can be treated as an
ordinary algebraic equation. For $n>2$ the loop equation above does
not determine $t(x)$.

For $n>2$ we need the higher loop equations. These are more difficult
to derive directly. However, we can deduce their form by noting
an interesting connection to the loop equations for the matrix model
associated to the $A_{n-1}$ quiver theory. (Here we mean the $\N=2$
quiver theory with polynomial superpotential deformations for the $n$
adjoint chiral multiplets. These softly break $\N=2$ supersymmetry to
$\N=1$.) The higher order loop
equations for the $A_{n-1}$ quiver matrix model were derived in
\cite{klemm} using free boson techniques. The quadratic $A_{n-1}$ loop
equations derived by them (see Eq.(2.64) of \cite{klemm}) is in fact
identical to our Eq.\eqref{loop1} provided we identify the $k$-th resolvent
of the quiver model $\omega_k(x)$ with $q^k\omega(q^k x)$ in our
theory. We need to make a similar identification of the deforming
polynomials. Basically, the different factors in the quiver are
constrained to be the same, up to a phase factor $q$ between adjacent
factors in the quiver. This is not a coincidence. It is related to the
fact that
the theory on D-branes at the orbifold $\C^3/\Z_n\times\Z_n$ with
discrete torsion can be obtained by the ``orbifolding of an orbifold'',
\cite{Douglas2} ({\it i.e.} ${\bf Z}_n$ orbifold of ${\bf C}\times{\bf
C}/{\bf Z}_n$). The phase factor $q$ between adjacent factors of the
quiver represents the single unit of discrete torsion. The $n$-th
order
loop equations for our theory can be obtained by applying this
prescription to the results of \cite{klemm}. It would also be
interesting to apply these ideas to the derivation of loop equations
from anomaly chains as done in \cite{Casero}.
\Appendix{Some Properties of Elliptic Functions}

In this appendix we provide some useful---but far from
complete---details of elliptic functions and their near cousins.
For definitions and a more complete
treatment, we refer the reader to one of the textbooks, for example
\cite{ww}. An elliptic function $f(u)$ is a function on the complex
plane, periodic in two periods $2\omega_1$ and $2\omega_2$. We will
define the lattice $\Gamma=2\omega_1{\mathbb Z}\oplus2\omega_2{\mathbb
Z}$ and define the basic period parallelogram as
\EQ{
{\cal
D}=\big\{u=2\mu\omega_1+2\nu\omega_2,\ 0\leq\mu<1,\ 0\leq\nu<1\big\}\ .
}
The complex structure of the torus defined by identifying the edges of
${\cal D}$ is
\EQ{
\tau=\omega_2/\omega_1
}
and we also define
\EQ{
q=e^{i\pi\tau}\ .
}

\subsection{The Weierstrass function}

The archetypal
elliptic function is the Weierstrass $\wp(u)$ function. It is an even
function which is
analytic throughout ${\cal D}$, except at $u=0$
where it has a double pole:
\SP{
&\wp(u)=\frac1{u^2}+\sum_{k=1}^\infty c_{k+1}u^{2k}\ ,\\
&c_2=\frac{g_2}{20}\ ,\quad c_3=\frac{g_3}{28}\ ,\quad
c_k=\frac{3}{(2k+1)(k-3)}\sum_{j=2}^{k-2}c_jc_{k-j}\quad k\geq4\ .
}
The Weierstrass function satisfies the fundamental identity
\EQ{
\Big(\frac{d}{du}\wp(u)\Big)^2=4\wp(u)^3-g_2\wp(u)-g_3
}
which defines the {\it Weierstrass invariants\/}
$g_{2,3}=g_{2,3}(\omega_\ell)$ associated
to the torus. If the values of $\wp(\omega_i)$ be $e_i$, $(i=1,2,3)$,
then $\{e_i\}$ are roots of the equation $4t^3-g_2t-g_3=0$. Here
$\omega_3=\omega_1+\omega_2$. From the formulae connecting roots of
equations with their coefficents, it follows that
\EQ{e_1+e_2+e_3=0,\quad e_1e_2+e_2e_3+e_3e_1=-{1\over 4}g_2,
\quad e_1e_2e_3 ={1\over 4}g_3.\label{winv}}
\subsection{The Weierstrass zeta function}

We are also interested in other functions which are only
quasi-elliptic. First we have
$\zeta(u)$. It is an odd function
with the quasi-elliptic property:
\EQ{
\zeta(u+2\omega_\ell)=\zeta(u)+2\zeta(\omega_\ell)\ .
}
Its derivative gives minus the Weierstrass function
\EQ{
\qquad\wp(u)=-\zeta'(u)\ .
}
It follows that
$\zeta(u)$ has a simple pole at $u=0$.
Useful identities are
\AL{
&\omega_2\zeta(\omega_1)-\omega_1\zeta(\omega_2)=\frac{\pi i}2\ ,\\
&\zeta(\omega_1+\omega_2)=\zeta(\omega_1)+\zeta(\omega_2).}
The (quasi)-elliptic functions evaluated at rational multiples of
their periods satisfy special relations. For example, one useful
identity that we need is
\EQ{\zeta(2\omega_i/3)-{2\over 3}\zeta(\omega_i)=\Big[{1\over 3}
\wp(2\omega_i/3)\Big]^{1/2}.}

\subsection{The Jacobi theta functions}

We are also interested in the Theta functions $\theta_i(x|\tau)$, or
$\theta_i(x,q)$, $i=1,2,3,4$. They are also quasi-elliptic
functions on ${\cal D}$ when $x=\pi u/2\omega_1$. Each of them satisfies
the heat equation
\EQ{
\pi i\frac{\partial^2\theta_i(x|\tau)}{\partial x^2}+
4\frac{\partial\theta_i(x|\tau)}{\partial\tau}=0\ .
\label{heat}
}
They are related to the
previous functions; for instance,
\EQ{
\zeta(u)-\frac{\zeta(\omega_1)}{\omega_1}u=\frac{\pi}{2\omega_1}
\left.\frac{\theta'_1(x|\tau)}{\theta_1(x|\tau)}\right|_{x=\pi
  u/2\omega_1}
\ ,
\label{relt}
}
where the derivative is with respect to $x$. By differentiating this
relation one can obtain further identities relating the Jacobi theta
functions to the Weierstrass function and its derivatives.

Another useful identity that we need:
\EQ{3\Big[{\theta_1^\prime(\pi/3|\tau)\over\theta_1(\pi/3|\tau)}\Big]^2
\Big({\pi\over 2\omega_1}\Big)^2=\wp(2\omega_1/3).}

\subsection{``$q$''-expansions}

To take classical limits (${\rm Im}(\tau)\rightarrow \infty$) we use
the following expansions
\EQ{\wp(u)=-{\zeta(\omega_1)\over\omega_1}+
\Big({\pi\over 2\omega_1}\Big)^2{\rm cosec}^2{\pi u\over 2\omega_1}
-2\Big({\pi\over\omega_1}\Big)^2\;{\sum_{n=1}^\infty}\;
{nq^{2n}\over 1-q^{2n}}{\rm cos}{n\pi u\over\omega_1}.}
Using this and $\wp(u)=-\zeta^\prime(u)$ and Eq.\eqref{relt}, the
$q$-expansion for $\theta_1^\prime/\theta_1$ can be easily derived.

For evaluating classical limits of $e_i$, the following expansion is
useful
\EQ{\wp(u)=\Big({\pi\over 2\omega_1}\Big)\Big[-{1\over 3}
+\sum_{n=-\infty}^\infty\;{\rm cosec}^2\Big({z-2n\omega_2\over
  2\omega_1}\pi\Big)-{\sum_{n=-\infty}^\infty}^\prime\;
{\rm cosec}^2{n\omega_2\over\omega_1}\pi \Big].}
\subsection{The expression of elliptic functions by means of
  Weierstrass functions:}
Here we quote the result in article 21.5 of \cite{ww} using
Eq.\eqref{relt} to obtain a slightly different form. Let $f(z)$ be an
elliptic function with a fundamental set ({\it i.e.} modulo
translations by a period) of zeros
$(\alpha_1,\alpha_2,\ldots\alpha_n)$ and poles
$(\beta_1,\beta_2,\ldots\beta_n)$ so that
\EQ{\sum_{r=1}^n(\alpha_r-\beta_r)=0.}
Then
\EQ{f(z)=A\prod_{r=1}^n \Big[\theta_1\Big({\pi z -\pi\alpha_r\over
    2\omega_1}|\tau\Big)/\theta_1\Big({\pi z -\pi\beta_r\over
    2\omega_1}|\tau\Big)\Big].}
where $A$ is a constant. Now if
\EQ{\sum_{m=1}^{m_r}A_{r,m}(z-\beta_r)^{-m}}
be the principal part of $f(z)$ at its pole $\beta_r$, then
\EQ{f(z)=A^\prime+\sum_{r=1}^{n}\Big\{-A_{r,1}\zeta(z-\beta_r)+\sum_{m=1}^{m_r-1}
{(-)^{m+1}A_{r,m+1}\over m!}{d^{m-1}\over
  dz^{m-1}}\wp(z-\beta_r)\Big\}\label{repweier}}
where $A^\prime$ is an appropriately chosen constant.


\begin{thebibliography}{99}


{\small
\bibitem{seiberg}
N.~Seiberg,
Nucl.\ Phys.\ B {\bf 435}, 129 (1995)
[arXiv:hep-th/9411149].


\bibitem{ks}
I.~R.~Klebanov and M.~J.~Strassler,
JHEP {\bf 0008}, 052 (2000)
[arXiv:hep-th/0007191].

\bibitem{Gubser:2004qj}
S.~S.~Gubser, C.~P.~Herzog and I.~R.~Klebanov,
[arXiv:hep-th/0405282].


\bibitem{ber}
D.~Berenstein,
JHEP {\bf 0111}, 060 (2001)
[arXiv:hep-th/0110184].

\bibitem{ber1}
D.~Berenstein,
Phys.\ Lett.\ B {\bf 552}, 255 (2003)
[arXiv:hep-th/0210183].


\bibitem{ber2}
D.~Berenstein,
JHEP {\bf 0306}, 019 (2003)
[arXiv:hep-th/0303033].

\bibitem{mm3}
N.~Dorey, T.~J.~Hollowood and S.~P.~Kumar,
JHEP {\bf 0212}, 003 (2002)
[arXiv:hep-th/0210239].


\bibitem{nickprem}
N.~Dorey and S.~P.~Kumar,
JHEP {\bf 0002}, 006 (2000)
{[arXiv:hep-th/0001103]}.


\bibitem{ADK}
O.~Aharony, N.~Dorey and S.~P.~Kumar,
JHEP {\bf 0006} (2000) 026 [arXiv:hep-th/0006008].


\bibitem{DV1}
R.~Dijkgraaf and C.~Vafa,
{[arXiv:hep-th/0208048]}

\bibitem{DV2}
R.~Dijkgraaf and C.~Vafa,
{[arXiv:hep-th/0207106]}

\bibitem{DV3}
R.~Dijkgraaf and C.~Vafa,
{[arXiv:hep-th/0206255]}



\bibitem{LS}
R.~G.~Leigh and M.~J.~Strassler,
Nucl.\ Phys.\ B {\bf 447}, 95 (1995) [arXiv:hep-th/9503121].

\bibitem{Douglas1}
M.~R.~Douglas,
[arXiv:hep-th/9807235].

\bibitem{Douglas2}
M.~R.~Douglas and B.~Fiol,
[arXiv:hep-th/9903031].


\bibitem{LB}
D.~Berenstein, V.~Jejjala and R.~G.~Leigh,
Nucl.\ Phys.\ B {\bf 589}, 196 (2000)
[arXiv:hep-th/0005087].

\bibitem{LB1}
D.~Berenstein, V.~Jejjala and R.~G.~Leigh,
Phys.\ Lett.\ B {\bf 493}, 162 (2000)
[arXiv:hep-th/0006168].

\bibitem{mn}
J.~M.~Maldacena and C.~Nunez,
Phys.\ Rev.\ Lett.\  {\bf 86}, 588 (2001)
[arXiv:hep-th/0008001].


\bibitem{civ}
F.~Cachazo, K.~A.~Intriligator and C.~Vafa,
Nucl.\ Phys.\ B {\bf 603}, 3 (2001)
[arXiv:hep-th/0103067]; F.~Cachazo, B.~Fiol, K.~A.~Intriligator,
S.~Katz and C.~Vafa,
Nucl.\ Phys.\ B {\bf 628}, 3 (2002)
[arXiv:hep-th/0110028].



\bibitem{mm1}
N.~Dorey, T.~J.~Hollowood, S.~Prem Kumar and A.~Sinkovics,
[arXiv:hep-th/0209089].

\bibitem{mm2}
N.~Dorey, T.~J.~Hollowood, S.~P.~Kumar and A.~Sinkovics,
[arXiv:hep-th/0209099].

\bibitem{kostov}
I.~K.~Kostov,
Nucl.\ Phys.\ B {\bf 575}, 513 (2000)
{[arXiv:hep-th/9911023]}.

\bibitem{dvdecon}
R.~Dijkgraaf and C.~Vafa,
[arXiv:hep-th/0302011].

\bibitem{KW} I.~R.~Klebanov and E.~Witten,
Nucl.\ Phys.\ B {\bf 536} 199 (1998) [arXiv:hep-th/9807080].

\bibitem{ami}A.~Hanany and J.~Walcher,
JHEP {\bf 0306}, 055 (2003) [arXiv:hep-th/0301231];
S.~Franco, A.~Hanany, Y.~H.~He and P.~Kazakopoulos,
[arXiv:hep-th/0306092]; S. Franco, A.~Hanany and Y.~H.~He,
Fortsch.\ Phys.\  {\bf 52}, 540 (2004) [arXiv:hep-th/0312222].

\bibitem{fiol}B.~Fiol,
JHEP {\bf 0207}, 058 (2002) [arXiv:hep-th/0205155].

\bibitem{PS}
J.~Polchinski and M.~J.~Strassler,
[arXiv:hep-th/0003136].


\bibitem{FK}
H. M. Farkas and I. Kra, ``Riemann Surfaces'', Springer-Verlag (1980).

\bibitem{glueball}
T.~J.~Hollowood,
JHEP {\bf 0310}, 051 (2003) [arXiv:hep-th/0305023].

\bibitem{Aganagic:2003qj}
M.~Aganagic, R.~Dijkgraaf, A.~Klemm, M.~Marino and C.~Vafa,
[arXiv:hep-th/0312085].














\bibitem{ww}
E. T. Whittaker, G. N. Watson, ``A course of Modern Analysis'', 4th edn.,
Cambridge University Press, 1927.



\bibitem{Kob} N.~Koblitz, ``Introduction to Elliptic Curves and
Modular Forms''. Springer-Verlag, 2nd Edition 1993.

\bibitem{klemm}
S.~Chiantese, A.~Klemm and I.~Runkel,
JHEP {\bf 0403}, 033 (2004)
[arXiv:hep-th/0311258].

\bibitem{Casero}
R.~Casero and E.~Trincherini,
JHEP {\bf 0309}, 041 (2003)
[arXiv:hep-th/0304123].

}

\end{thebibliography}
\end{document}